\documentclass[showpacs,amsmath,amssymb,aps,showkeys,floatfix,prd,a4paper]{revtex4}

\usepackage[dvips]{graphicx}
\usepackage{dcolumn}
\usepackage{bm}
\usepackage{epsfig}
\usepackage{amsfonts}
\usepackage{amssymb,amscd}

\def\lsim{\raise0.3ex\hbox{$<$\kern-0.75em\raise-1.1ex\hbox{$\sim$}}}
\def\gsim{\raise0.3ex\hbox{$>$\kern-0.75em\raise-1.1ex\hbox{$\sim$}}}

\def\beq{\begin{equation}}
\def\eeq{\end{equation}}
\def\bea{\begin{eqnarray}}
\def\eea{\end{eqnarray}}
\def\bq{\begin{quote}}
\def\eq{\end{quote}}

\newcommand{\rr}{\mbox{\boldmath $r$}}

\def\gappeq{\mathrel{\rlap {\raise.5ex\hbox{$>$}}
{\lower.5ex\hbox{$\sim$}}}}

\def\lappeq{\mathrel{\rlap{\raise.5ex\hbox{$<$}}
{\lower.5ex\hbox{$\sim$}}}}

\def\Toprel#1\over#2{\mathrel{\mathop{#2}\limits^{#1}}}

\begin{document}


\title{Color dipole predictions for the exclusive vector meson 
photoproduction in $pp/pPb/PbPb$ collisions at Run 2 LHC energies }

\author{V.~P. Gon\c{c}alves $^1$, M.V.T. Machado $^2$, B. D. Moreira $^1$, 
F. S. Navarra $^{3,4}$, G. Sampaio dos Santos $^1$ }
\affiliation{$^1$ High and Medium Energy Group, \\
Instituto de F\'{\i}sica e Matem\'atica, Universidade Federal de Pelotas\\
Caixa Postal 354, CEP 96010-900, Pelotas, RS, Brazil \\ 
$^2$ High Energy Physics Phenomenology Group, GFPAE  IF-UFRGS \\
Caixa Postal 15051, CEP 91501-970, Porto Alegre, RS, Brazil  \\
$^3$ Instituto de F\'{\i}sica, Universidade de S\~{a}o Paulo,
C.P. 66318,  05315-970 S\~{a}o Paulo, SP, Brazil\\
$^4$ Institut de Physique Th\'eorique, Universit\'e Paris Saclay,\\
CEA, CNRS, F-91191, Gif-sur-Yvette, France }


\date{\today}

\begin{abstract}
In this paper we present a comprehensive analysis of  exclusive vector meson photoproduction in $pp$, $pPb$ and $PbPb$ collisions at Run 2 LHC energies using the Color Dipole formalism. The rapidity distributions and total cross sections for the $\rho$, $\phi$, $J/\Psi$, $\Psi (2S)$ and $\Upsilon$ production are estimated considering the more recent phenomenological models for the dipole - proton scattering amplitude, which are based on the Color Glass Condensate formalism and are able to describe the inclusive and exclusive $ep$ HERA data. Moreover, we also discuss the impact of the modelling of the vector meson wave functions on the predictions. The current theoretical uncertainty in the Color Dipole predictions is estimated  and a comparison  with the experimental results is performed. 

\end{abstract}

\maketitle

\section{Introduction}
\label{intro}

One of the goals of  hadron physics is to achieve a deeper knowledge of the 
hadronic structure. In particular, the advent of the high energy colliders  has 
motivated the study of the hadron structure at high energies. 
An important phenomenological and experimental tool for this purpose is the deep 
inelastic scattering (DIS) $ep$, where an electron emits a virtual photon which 
interacts with a proton target. The proton structure can then be studied through 
the $\gamma^{*}p$ interaction, taking into account the QCD dynamics at high energies. 
The experimental study of DIS was carried out at HERA, where the  $\gamma^{*}p$ 
c.m. energy ($W_{max}$) reached a maximum value  of the order of 200 GeV.  
HERA data have shown that the gluon density inside the proton grows with the energy.
Therefore, at high energies, a hadron becomes a dense system and  the nonlinear effects 
inherent to  the QCD dynamics may become visible. This dense system is best described by 
the Color Glass Condensate (CGC) approach \cite{hdqcd}.   
In the future an electron-ion collider may be built \cite{eics}. This will be the ideal machine for 
the study of  hadron structure and of QCD dynamics, specially with heavy nuclei. This will also be the best place to 
test the CGC formalism for DIS.  

Alternatively, one can study  the $\gamma p(A)$ interaction at the LHC, in 
ultraperipheral collisions  (UPC) and reach higher energies ($W \sim 900 - 8000$ GeV) 
than those reached at HERA. In a UPC at high energies, two charged hadrons (or nuclei)
interact at impact parameters larger than the sum of their radii \cite{upc}. Under these 
circunstances, it is well known that the hadron acts as a source of almost real photons 
and  photon-photon or photon-hadron interactions may happen.   The first part of the 
process, the photon emission, is a pure QED process while, in the second part, the 
photon-photon or photon-hadron interaction, other interactions may take place. In this 
work we study the QCD dynamics in exclusive vector meson photoproduction in photon -- hadron interactions. 

The first studies of  exclusive vector meson photoproduction in UPC's were made in Refs. \cite{klein_prc,gluon,strikman}. Since then several theoretical works related to 
this subject have been published \cite{outros_klein,vicmag_mesons1,
outros_vicmag_mesons,outros_frankfurt,Schafer,vicmag_update,gluon2,motyka_watt,
Lappi,griep,Guzey,Martin,glauber1,bruno1,Xie,bruno3,vicnavdiego,tuchin}. On the experimental side, a great amount of data has 
been accumulated 
\cite{cdf,star,phenix,alice,alice2,lhcb,lhcb2,lhcb3,lhcbconf,review_forward}. 
During the  
last years, the LHC has provided data on vector meson photoproduction at 
Run 1 energies and in this year at Run 2 energies.  The Run 2 at the LHC has already 
produced $PbPb$ collisions and  more data in $pp/pPb/PbPb$ collisions are expected in the next years. These collisions are now  performed at energies which are a factor $\approx 2$ larger than those of Run 1. The measurements at central rapidities will be sensitive to larger
$\gamma Pb$ energies and hence to smaller $x$ values of the nuclear gluon distribution. Similarly, the experimental results obtained in $pp$ and $pPb$ collisions will probe the QCD dynamics at small - $x$.
The study of  vector meson ultraperipheral photoproduction has  thus  high 
priority. 

Given that the QED part of the ultraperipheral interaction is well known, we can use 
this process to constrain the QCD dynamics. In Ref. \cite{vicmag_mesons1} the authors 
proposed to study  exclusive vector meson photoproduction in ultraperipheral collisions using the color dipole picture. In this formalism  the photon fluctuates into a $q\bar{q}$ which interacts with the hadron target via strong interaction and then turns into a vector meson. The main ingredients for the calculation of the cross sections are the vector meson wave function and the dipole - hadron scattering amplitude, which is dependent on the modelling of the QCD dynamics at high energies.       
Over the last years several authors \cite{outros_vicmag_mesons,vicmag_update,motyka_watt,
Lappi,griep,glauber1,bruno1,Xie,bruno3,vicnavdiego,tuchin} 
have studied exclusive  vector meson photoproduction in $pp/pPb/PbPb$ collisions using the Color Dipole formalism, considering different models to describe the dipole-target  interaction and distinct approaches to treat the vector meson wave functions. As a consequence, a direct comparison between the predictions obtained in different studies is not an easy task. Our goal in this paper is to perform a comprehensive analysis of  exclusive  vector meson photoproduction in hadronic collisions considering the three phenomenological models for the dipole -- proton scattering amplitude, which are able to describe the  high precision HERA data on inclusive and exclusive $ep$ processes, as well as two models for the vector meson wave function. Such analysis allows to estimate the current theoretical uncertainty in the color dipole predictions. The comparison with the experimental data will allow to constrain the modelling of the QCD dynamics and of the vector meson function. Moreover, it
will tell us how much additional ingredients, as e.g. the inclusion of a survival gap factor \cite{Schafer,Martin}, next - to - leading order corrections \cite{wallon}  and/or nuclear shadowing \cite{Guzey,glauber1}, are necessary to describe  vector meson photoproduction in hadronic collisions. Finally, our goal is to provide, for the first time, predictions for  exclusive light vector photoproduction in $pp/pPb/PbPb$ collisions at the Run 2 energies and to complement the predictions presented in Ref. \cite{bruno3}. In that work,  exclusive heavy vector meson photophotoproduction in $pp/PbPb$ collisions was treated with the impact parameter Color Glass Condensate (bCGC) model for the dipole - proton scattering amplitude $({\cal{N}})$ and the Boosted - Gaussian model for the vector meson wave function ($\Psi^V$). Here we also present predictions considering the IIM and IP-SAT models for ${\cal{N}}$ and the Gaus-LC model for $\Psi^V$. In particular, the results obtained the updated version of the IP-SAT model are presented here for the first time.

The paper is organized as follows. In  Sec. \ref{formalism} we present a brief review of the Color Dipole formalism and the main expressions 
used to estimate   exclusive photoproduction of vector mesons. Moreover, the distinct models for the dipole scattering amplitude and  vector meson wave function are discussed.  In Sec. \ref{res}, we present our predictions for the cross 
sections and 
rapidity distributions to be measured in $\rho$, $\phi$, $J/\Psi$, $\Psi (2S)$ and $\Upsilon$ production in $pp/pPb/PbPb$ collisions at the Run 2 LHC energies. 
Finally, in Sec. \ref{conc}, we summarize our main conclusions.

\section{Formalism}

\label{formalism}

In this section we will present a brief review of the formalism. Let us start 
defining a UPC as a collision between two electric charges at impact 
parameters such that $b > R_{1} + R_{2}$, where $R_{i}$ is the radius of the charge 
$i$. In a UPC at high energies, it is well known that the hadrons act as a source of 
almost real photons and the hadron-hadron cross section can be written 
in a factorized form, 
given by the so called equivalent photon approximation (Ref. \cite{upc})
\begin{eqnarray}
 \sigma(h_{1} + h_{2} \rightarrow h_{1} \otimes V \otimes h_{2}) &=& 
\int d\omega \frac{n_{h_{1}}(\omega)}{\omega} 
\sigma_{\gamma h_{2} \rightarrow V\otimes h_{2}}\left( W_{\gamma h_{2}}^{2}\right) 
+ \int d\omega \frac{n_{h_{2}}(\omega)}{\omega} 
\sigma_{\gamma h_{1} \rightarrow V\otimes h_{1}}\left( W_{\gamma h_{1}}^{2}\right)  .
\label{epa}
\end{eqnarray}
In this equation, $\otimes$ represents the presence of a rapidity gap in the final state, $n(\omega)$ is the equivalent photon spectrum generated by  the 
hadronic source and $\sigma_{\gamma h \rightarrow V \otimes h }( W_{\gamma h}^{2})$ 
is the vector meson photoproduction cross section. Moreover, $\omega$ and 
$W_{\gamma h}$  are  the photon energy and the c.m. energy of the $\gamma h$ system, 
where 
\begin{eqnarray}
 W_{\gamma h} = \sqrt{4 \omega E} , \,\,\, E = \sqrt{s}/2 \,\,\, 
\end{eqnarray}
and $\sqrt{s}$ is the hadron-hadron c.m. energy.
The equivalent photon spectrum is fully computed in QED. For the case where a 
nucleus is the source of photons, we have \cite{upc}
 \begin{eqnarray}
   n_{A}(\omega) = \frac{2Z^{2}\alpha_{em}}{\pi }  
\left[
\xi K_{0}(\xi) K_{1}(\xi) -\frac{\xi^{2}}{2} \left( K_{1}^{2}(\xi) - K_{0}^{2}(\xi)  
\right )  \right]  ,
  \end{eqnarray}
where
\begin{eqnarray}
 \xi = \omega \left( R_{h_{1}} + R_{h_{2}} \right) / \gamma_{L} ,
\end{eqnarray}
with $\gamma_L$ being the target frame Lorentz boost. On the other hand, if a proton is the source of photons the spectrum can be approximated by \cite{Dress}
\begin{eqnarray}
  n_{p}(\omega) = \frac{\alpha_{em}}{2 \pi }  
 \left[ 1 + \left(  1 - \frac{2\omega}{\sqrt{s}}        \right)^{2}     \right] 
\left(
\ln \Omega - \frac{11}{6} + \frac{3}{\Omega}  -   \frac{3}{2\Omega^{2}} + 
\frac{1}{3\Omega^{3}}  \right) ,
\end{eqnarray}
where
\begin{eqnarray}
 \Omega = 1 + [
(0.71\,\mbox{GeV}^{2}) / Q_{min}^{2}
]
\end{eqnarray}
and
\begin{eqnarray}
 Q_{min}^{2} = \omega^{2} / [\gamma_{L}^{2}(1-2\omega/\sqrt{s})] 
\approx (\omega / \gamma_{L})^{2}.
\end{eqnarray}

In order to estimate the  exclusive vector meson photoproduction in hadronic collisions using  Eq.(\ref{epa}) we need to know the $\gamma \, h \to V \, h$ cross section.
The cross section for exclusive vector meson production 
is given by 
\begin{eqnarray}
 \sigma(\gamma h \rightarrow V h) = \int_{-\infty}^{0} dt \,\,\frac{d \sigma}{dt} 
= \frac{1}{16 \pi}  \int_{-\infty}^{0} 
\left |  
{\cal A}^{\gamma h \rightarrow V h} (x, \Delta)  
\right|^{2}   dt.  
\label{cs_gammap}
\end{eqnarray}
The scattering amplitude ${\cal A}^{\gamma h \rightarrow V h} (x, \Delta)$ will be derived using the color dipole formalism \cite{nik}, which allows us to study  the $\gamma h$ interaction in terms of a  (color) dipole - hadron interaction. 
In this formalism,  
we assume that the photon fluctuates into a color dipole which interacts with the 
hadron target and then forms a vector meson at the final state. If the lifetime of  
the dipole is much larger than the  interaction time, a condition which is satisfied
in high energy collisions, the quasi - elastic scattering amplitude for the process 
$\gamma h \rightarrow V h$  can be written as \cite{vicmag_mesons,KMW}
\begin{eqnarray}
 {\cal A}^{\gamma h \rightarrow V h} (x, \Delta) =  
i \int dz \,\, d^{2} \mbox{\textbf{\textit{r}}} \,\, d^{2} 
\mbox{\textbf{\textit{b}}}_{h}                          
[\Psi^{V*}\left( \mbox{\textbf{\textit{r}}}, z \right)   
\Psi\left(\mbox{\textbf{\textit{r}}}, z \right)]_T \,\,
e^{-i\left[\mbox{\textbf{\textit{b}}}_{h} - (1-z) \mbox{\textbf{\textit{r}}}        
\right]  \cdot \mathbf{\Delta} }
\,\,2\,\, {\cal N}_{h}(x,\mbox{\textbf{\textit{r}}},\mbox{\textbf{\textit{b}}}_{h}) \,\,, 
\end{eqnarray}
where the function $[\Psi^{V*} \left( \mbox{\textbf{\textit{r}}}, z \right) 
\Psi\left( \mbox{\textbf{\textit{r}}}, z \right)]_T$ is the overlap between the 
wave functions of the transverse photon and of the vector meson, which describes the fluctuation 
of the photon with transverse polarization into a color dipole and the subsequent formation of the 
vector meson. Furthermore, ${\cal N}_{h}(x,\mbox{\textbf{\textit{r}}},
\mbox{\textbf{\textit{b}}}_{h})$ is the imaginary part of the forward dipole - hadron 
scattering amplitude and it  carries all the information about the strong interactions 
in the process. The variables $z,\mbox{\textbf{\textit{r}}} , 
\mbox{\textbf{\textit{b}}}_{h} $ are, respectively, the light cone longitudinal momentum 
fraction of the photon carried by the quark (and $1-z$, for the antiquark), 
the transverse separation of the color dipole and the impact parameter, the separation
between the dipole center  and the target center. Further,  $x$ ($=M_{V}^{2}/W^{2}$) is the Bjorken 
variable for a diffractive event  and $ \mathbf{\Delta} $ is the 
Fourier conjugate of  $\mbox{\textbf{\textit{b}}}_{h}$. It is  related to the 
momentum transfer squared  by $\Delta =  \sqrt{ - t}$.

\begin{figure}[!t]
\begin{tabular}{cc}
\centerline{{\includegraphics[height=6cm]{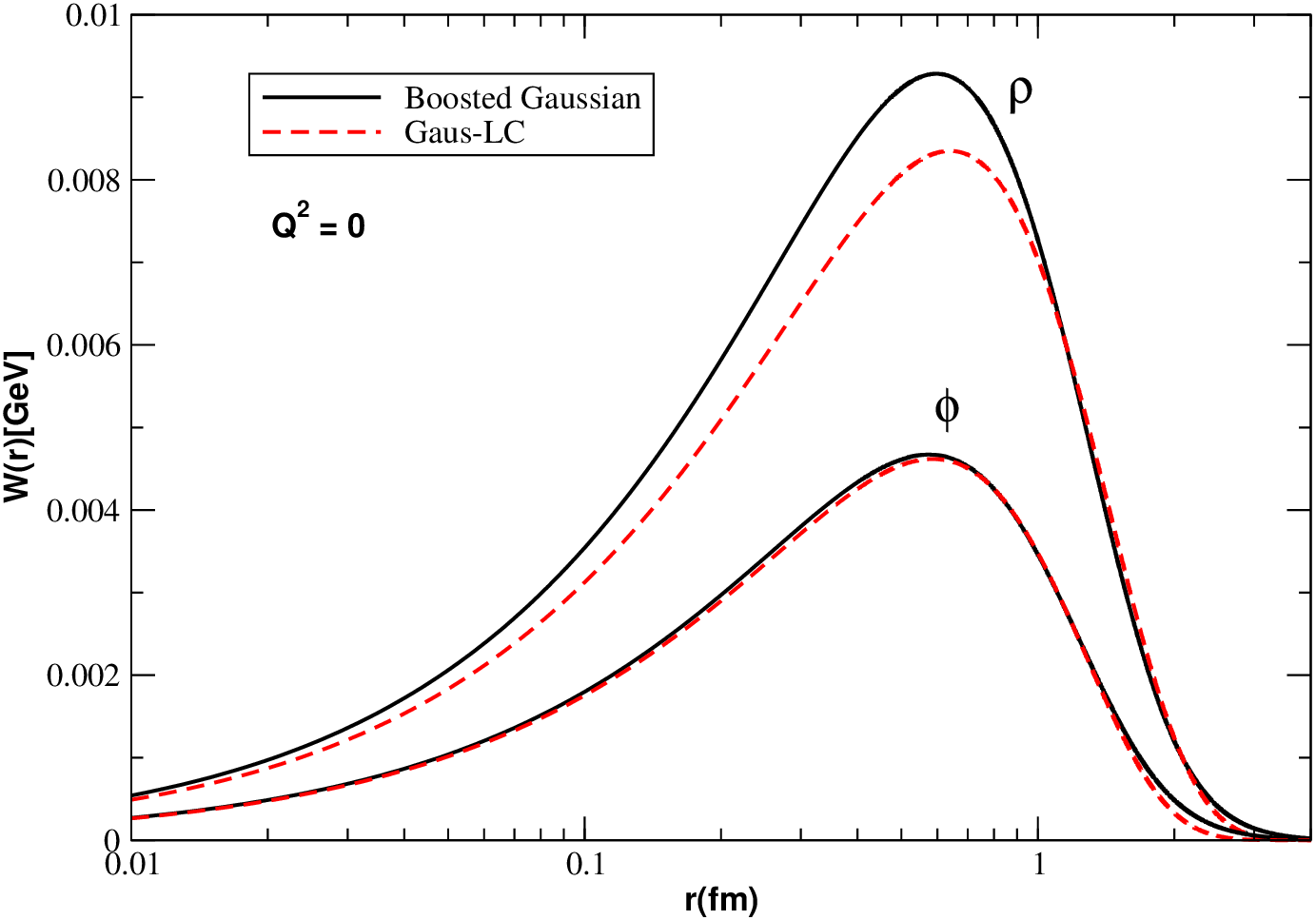}}
{\includegraphics[height=6cm]{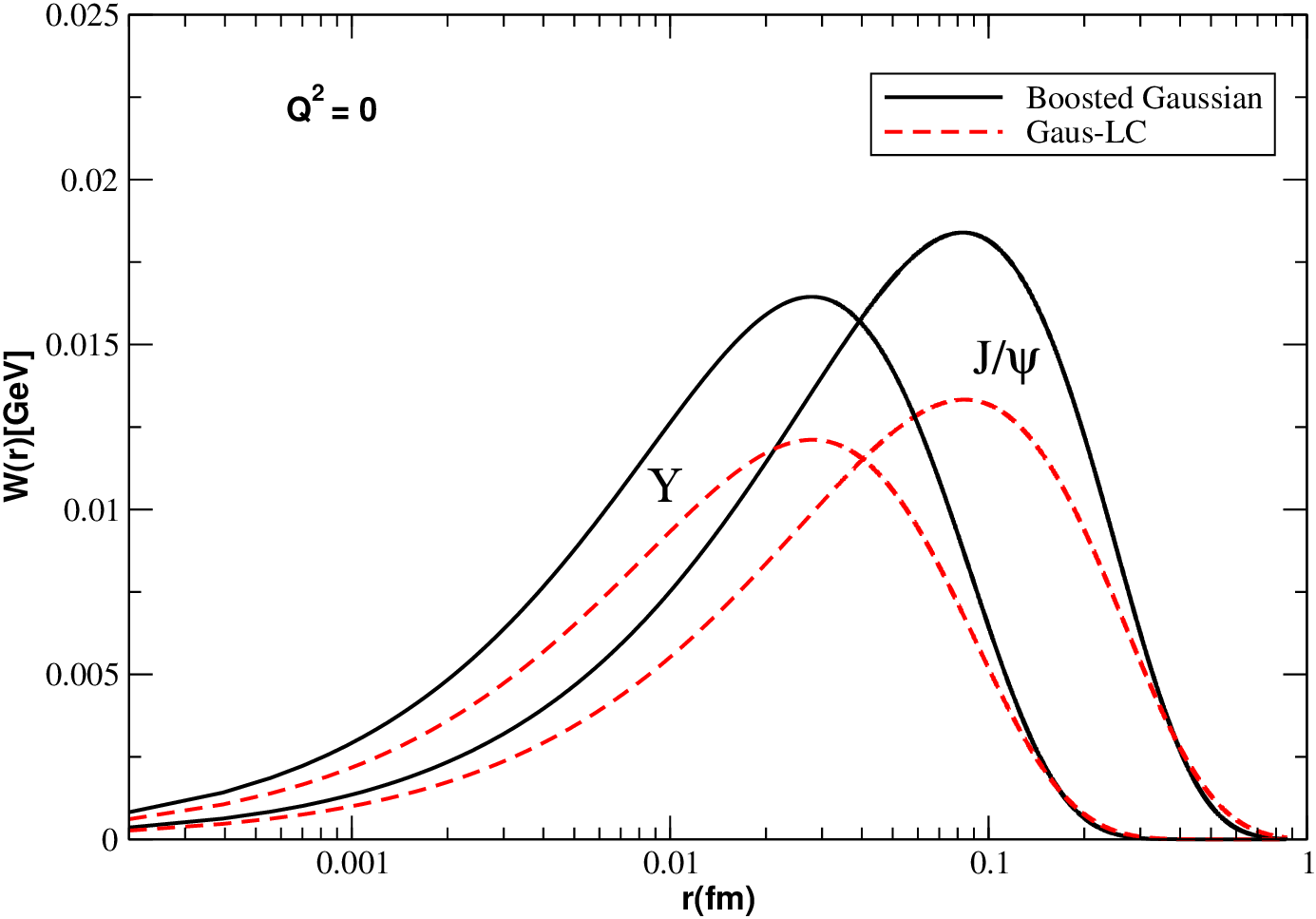}}}
\end{tabular}
\caption{Dependence on the dipole size $\rr$ of the function $ W(\mbox{\textbf{\textit{r}}})$, defined in Eq. \ref{overlap}, for light (left panel) and heavy (right panel) vector mesons, obtained assuming a real photon ($Q^2 = 0$) in the initial state.}
\label{wf}
\end{figure}

In the Color Dipole formalism the main ingredients for the calculation of the vector meson cross sections are the vector meson wave  function $\Psi^{V}$ and the dipole - target scattering amplitude ${\cal N}$. The treatment of both quantities is 
the subject of intense study by several groups. In particular, the description of the vector meson wave function is still a theme of debate, with different models being able to describe e.g. the HERA data. In what follows, we will consider two popular models in the literature: the Boosted Gaussian and the Gaus-LC models (For alternative descriptions see. e.g. Refs. \cite{forshaw,Li}). In these models the vector meson  is assumed to be predominantly a quark-antiquark state.
It is also assumed that the spin and polarization structure is the same as in the  photon \cite{wflcg,wfbg,sandapen,ipsat2}. As a consequence, the overlap between the photon and the vector meson wave function, for the transversely polarized  
case, is given by (For details see Ref. \cite{KMW})
\begin{eqnarray}
(\Psi_{V}^* \Psi)_T = \hat{e}_f e \frac{N_c}{\pi z (1-z)}\left\{m_f^2K_0(\epsilon r)\phi_T(r,z) -[z^2+(1-z)^2]\epsilon K_1(\epsilon r) \partial_r \phi_T(r,z)\right\} \,\,,
\end{eqnarray}
where $ \hat{e}_f $ is the effective charge of the vector meson, $m_f$ is the quark mass, $N_c = 3$, $\epsilon^2 = z(1-z)Q^2 + m_f^2$ and $\phi_T(r,z)$ define the scalar part of the  vector meson wave function. The Boosted Gaussian and Gaus-LC models differ in the assumption about the function $\phi_T(r,z)$. 
In the Boosted Gaussian model the function $\phi_T(r,z)$ is given by
\begin{eqnarray}
\phi_T(r,z) = N_T z(1-z) \exp\left(-\frac{m_fR^2}{8z(1-z)} - \frac{2z(1-z)r^2}{R^2} + \frac{m_f^2R^2}{2}\right) \,\,.
\end{eqnarray}
 In contrast, in the Gaus-LC model, it is given by
\begin{eqnarray}
\phi_T(r,z) = N_T [z(1-z)]^2 \exp\left(-\frac{r^2}{2R_T^2}\right)
\end{eqnarray}
The parameters $N_T$, $R$ and $R_T$ are  determined by the normalization condition of the wave function and by the decay width (See e.g. Refs. \cite{KMW,glauber1,bruno1,armesto_amir} for details). In  Fig. \ref{wf}, we present the behavior 
of the quantity
\begin{eqnarray}
  W(\mbox{\textbf{\textit{r}}}) = 2 \pi r \int_{0}^{1} dz \left[ 
\Psi^{V*}(\mbox{\textbf{\textit{r}}},z) 
  \Psi(\mbox{\textbf{\textit{r}}},z) \right],
 \label{overlap} 
\end{eqnarray}
as a function of  $\rr$ for different vector mesons, considering 
these two wave function models. 
We find that both models predict a peak in the function $W(\mbox{\textbf{\textit{r}}})$. The position of the peak is almost model independent, with the normalization of the Gauss-LC model being smaller than the Boosted Gaussian
one. Moreover,  the peak occurs at larger values of $\rr$ for light mesons. We can observe that the heavier mesons 
are associated to smaller dipoles.  
So, studying different 
final states we are mapping different configurations of the dipole size, which 
probe different regimes of the QCD dynamics.
This indicates that the  global analysis of these different final states is ideal to constrain the description of the high energy regime of the strong interactions.

\begin{figure}[!t]
\begin{tabular}{cc}
\centerline{
{
\includegraphics[height=6cm]{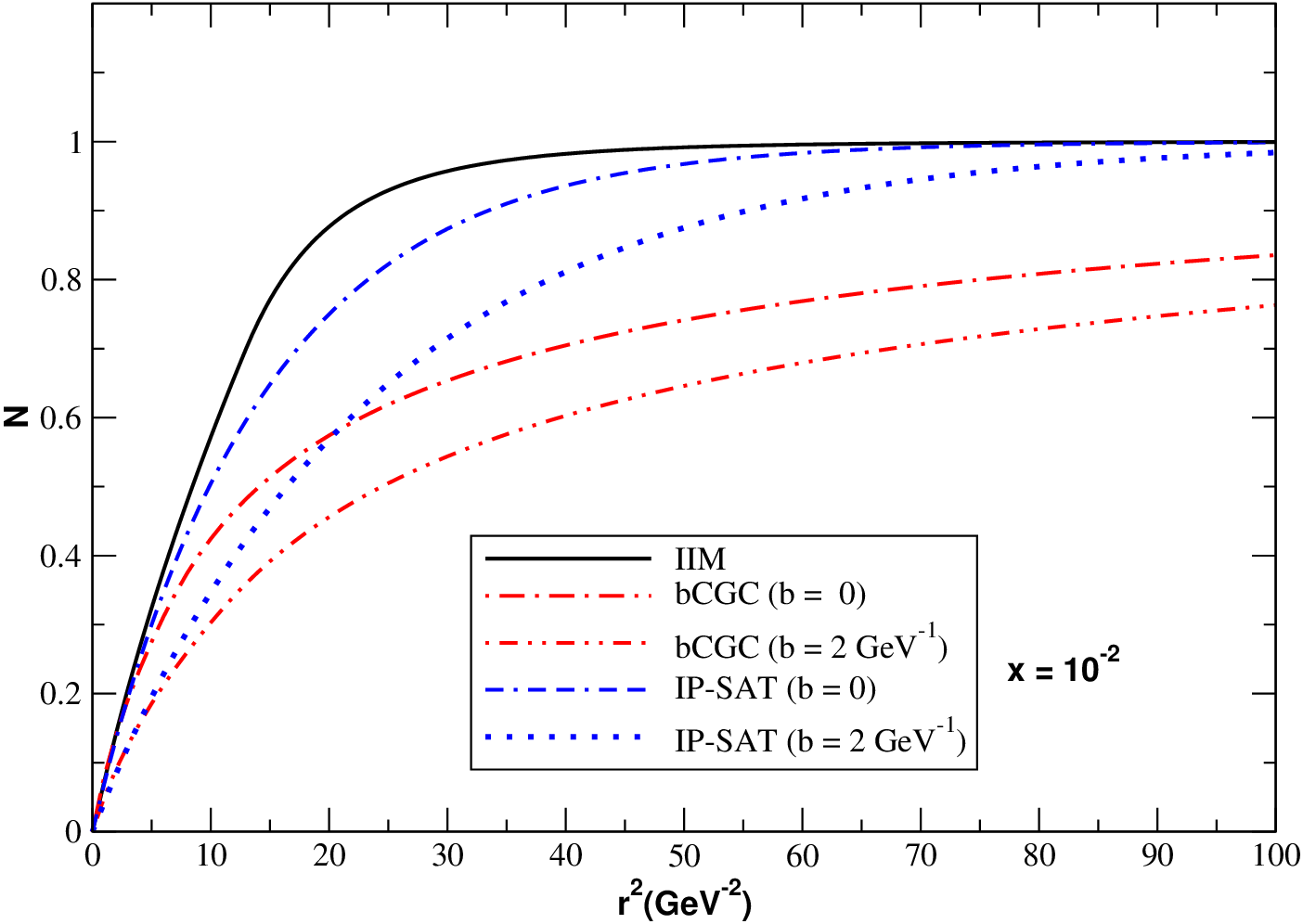}
}
{
\includegraphics[height=6cm]{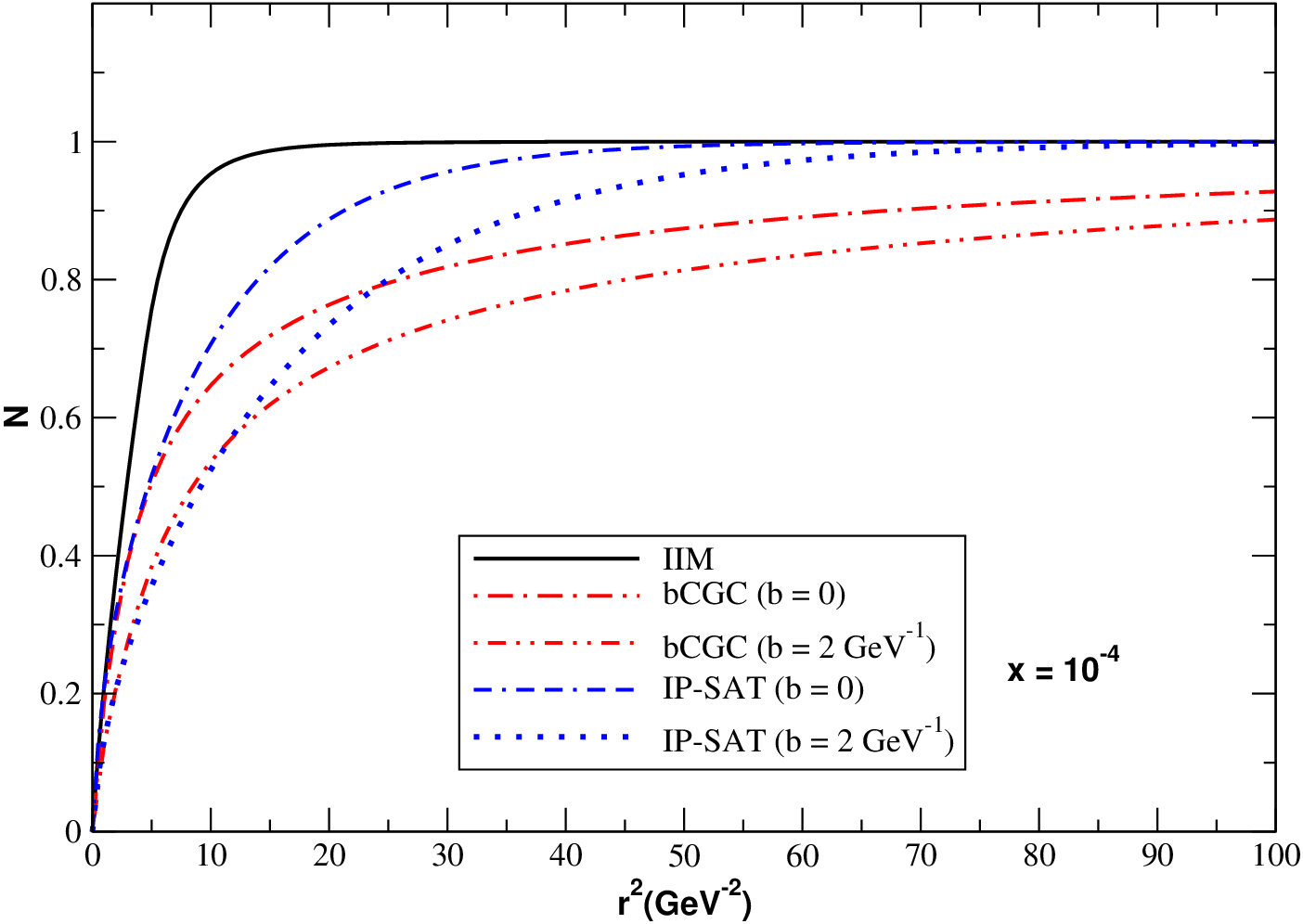}
}}
\end{tabular}
\caption{Dependence of dipole - proton scattering amplitude ${\cal N}_{p}$ as a function of $r^2$ for fixed values of $b$ and different values of $x$: $x=10^{-2}$ (left panel) and  $x=10^{-4}$ (right panel).}
\label{Np}
\end{figure}

Now let us  discuss some features of the models for the dipole - hadron scattering amplitude  ${\cal N}_{h}$. In the case of a proton target, the color dipole 
formalism has been extensively used to describe the inclusive and exclusive HERA data. During the last decades, several phenomenological models based on the Color Glass Condensate formalism have been proposed to describe the HERA data taking into account the nonlinear effects in the QCD dynamics. In general, such models differ in the treatment of the impact parameter dependence and/or of the linear and nonlinear regimes. Three examples of very successful models are the  IIM \cite{IIM_plb}, bCGC \cite{Watt_bcgc,KMW} and IPSAT \cite{ipsat1} models, which  have been updated in Refs. \cite{Rezaeian_update,ipsat4} using the high precision HERA data to constrain their 
free parameters and describe the data quite well. As we will estimate the vector meson photoproduction in hadronic collisions considering these three models, let
us  present a brief review of their main aspects. Initially, let us 
consider the Iancu -- Itakura -- Munier (IIM) model (also called CGC model) proposed in Ref. \cite{IIM_plb}.   
This model interpolates two analytical 
solutions of well known evolution equations: the solution of the BFKL equation near  the 
saturation regime and the solution of the  Balitski-Kovchegov 
equation deeply inside the saturation regime. The IIM model assumes that the impact parameter dependence of the scattering amplitude can be factorized as follows  ${\cal N}_{p}(x,\mbox{\textbf{\textit{r}}},
\mbox{\textbf{\textit{b}}}_{p}) = {\cal N}_{p}(x,\mbox{\textbf{\textit{r}}}) \, S(
\mbox{\textbf{\textit{b}}}_{p})$, where $S(
\mbox{\textbf{\textit{b}}}_{p})$ is the proton profile function and the dipole - proton scattering amplitude  ${\cal N}_{p}(x,\mbox{\textbf{\textit{r}}}) $ is given by \cite{IIM_plb}
\begin{eqnarray}
{\cal N}_p(x,r) = \left\{  \begin{array}{l}
{\cal N}_{0}\left( 
\frac{rQ_{s}}{2}
\right )^{2[\gamma_{s}+(1/(\kappa \lambda Y))\ln (2/rQ_{s})]} 
\,\,\, , rQ_{s} \leq 2\\
1- e^{-A \ln^{2}(BrQ_{s})} \,\,\,\,\,\,\,\,\,\,\,\,\,\,\,\,\,\,\,\,\,\,\,\,\,\,\,\,\,\,\,\,\,\,\,\,\, , rQ_{s} > 2
\end{array} \right.  
\label{iim}
\end{eqnarray}
where  $Y=\ln(1/x)$  and
\begin{eqnarray}
 Q_{s}(x) = \left(
\frac{x_{0}}{x}
 \right )^{\lambda/2}  ,
 \label{qs_iim}
\end{eqnarray}
is the saturation scale of this model. 
The free parameters were fixed by fitting the HERA data. Here we have used the updated 
parameters from Ref. \cite{Rezaeian_update}. The coefficients $A$ and $B$ are determined 
by the continuity condition of ${\cal N}$ and its derivative and are given by
\begin{eqnarray}
 A &=& - \frac{{\cal N}_{0}^{\,2}\gamma_{s}^{2}}
{(1-{\cal N}_{0})^{2} \ln (1-{\cal N}_{0})} \,\,\, , \\
B &=& \frac{1}{2}(1-{\cal N}_{0})^{-(1-{\cal N}_{0})/({\cal N}_{0}\gamma_{s})}  .
\end{eqnarray}
In Refs. \cite{KMW,Watt_bcgc}  a modification of this model was proposed, in 
order to include the impact parameter dependence. There, the authors have 
called this 
parametrization  bCGC model. The functional form of  ${\cal N}_p$ is the same as in 
Eq.(\ref{iim}), but  the saturation scale has the following dependence on $b$
\begin{eqnarray}
  Q_{s} \equiv Q_{s}(x,b) = 
\left(
\frac{x_{0}}{x}
 \right )^{\lambda/2} \left[
\exp \left(
-\frac{b^{2}}{2B_{CGC}}
 \right )
 \right ]^{1/(2 \gamma_{s})}  .
\label{qs_bcgc}
\end{eqnarray}
The parameters used are from Ref. \cite{Rezaeian_update}. Finally, the last model used 
was the IP-SAT model \cite{ipsat1,ipsat2,ipsat3}. This model uses an eikonalized 
form for 
${\cal N}_p$  that depends on a gluon distribution evolved via DGLAP equation and is given 
by 
\begin{eqnarray}
 {\cal N}_p(x,\mbox{\textbf{\textit{r}}},\mbox{\textbf{\textit{b}}}) = 
 1 - \exp \left[
\frac{\pi^{2}r^{2}}{N_{c}} \alpha_{s}(\mu^{2}) \,\,xg\left(x, \frac{4}{r^{2}} + 
\mu_{0}^{2}\right)\,\, T_{G}(b) 
 \right] ,
 \label{ipsat}
\end{eqnarray}
with a  Gaussian profile
\begin{eqnarray}
T_{G}(b) = \frac{1}{2\pi B_{G}}  
\exp\left(-\frac{b^{2}}{2B_{G}} \right) .
\end{eqnarray}
The initial gluon distribution evaluated at $\mu_{0}^{2}$ is taken to be 
\begin{eqnarray}
xg(x,\mu_{0}^{2}) =  A_{g}x^{-\lambda_{g}} (1-x)^{5.6} .
\end{eqnarray}
The free parameters of this model are fixed by a fit of HERA data. In this work we 
have used  a FORTRAN library (provided by the authors of Ref. \cite{ipsat4}) to  
calculate ${\cal N}$, which includes an updated analysis of combined HERA data.
In Fig. \ref{Np} we present a comparison between the IIM, bCGC and IP-SAT predictions for the  dipole-proton scattering amplitude 
as a function of  $r^{2}$ for two different values of the Bjorken variable $x$. 
For the $b$-dependent models, we show the results for two different values 
of $b$.  As it can be seen, the predictions of the different models are very 
different at large - $x$, with the differences decreasing at smaller values of $x$.
For small dipole sizes, we can observe the different $r$ dependence of the 
distinct models. In the IIM and the bCGC models, we observe that 
${\cal N} \propto \rr^{2 \gamma_{eff}}$ for $r^2 \rightarrow 0$, with different values for $\gamma_{eff}$. In contrast, the IP - SAT model predicts that ${\cal N} \propto \rr^{2} \, xg(x,4/r^2)$ in this limit.  On the other hand, for large dipole sizes,  the IIM and IP-SAT amplitudes have the same asymptotic value  while the bCGC amplitude  goes to a somewhat smaller value. The main difference between the models is associated to the behavior predicted for the transition between the linear (small - $r^{2}$)  and nonlinear (large - $r^{2}$) regimes of the QCD dynamics.  The 
IIM model predicts a more rapid transition than  the those predicted by the  
two $b$-dependent models. It is important to note that these three models for the dipole scattering amplitude describe the inclusive and exclusive HERA data. 
Since the production of the different vector mesons probes distinct values 
of $r$, as observed in Fig. \ref{wf}, their analysis can be useful to 
discriminate between the different models for the dipole - proton 
scattering amplitude.

\begin{figure}[!t]
\begin{tabular}{cc}
\centerline{
{
\includegraphics[height=6cm]{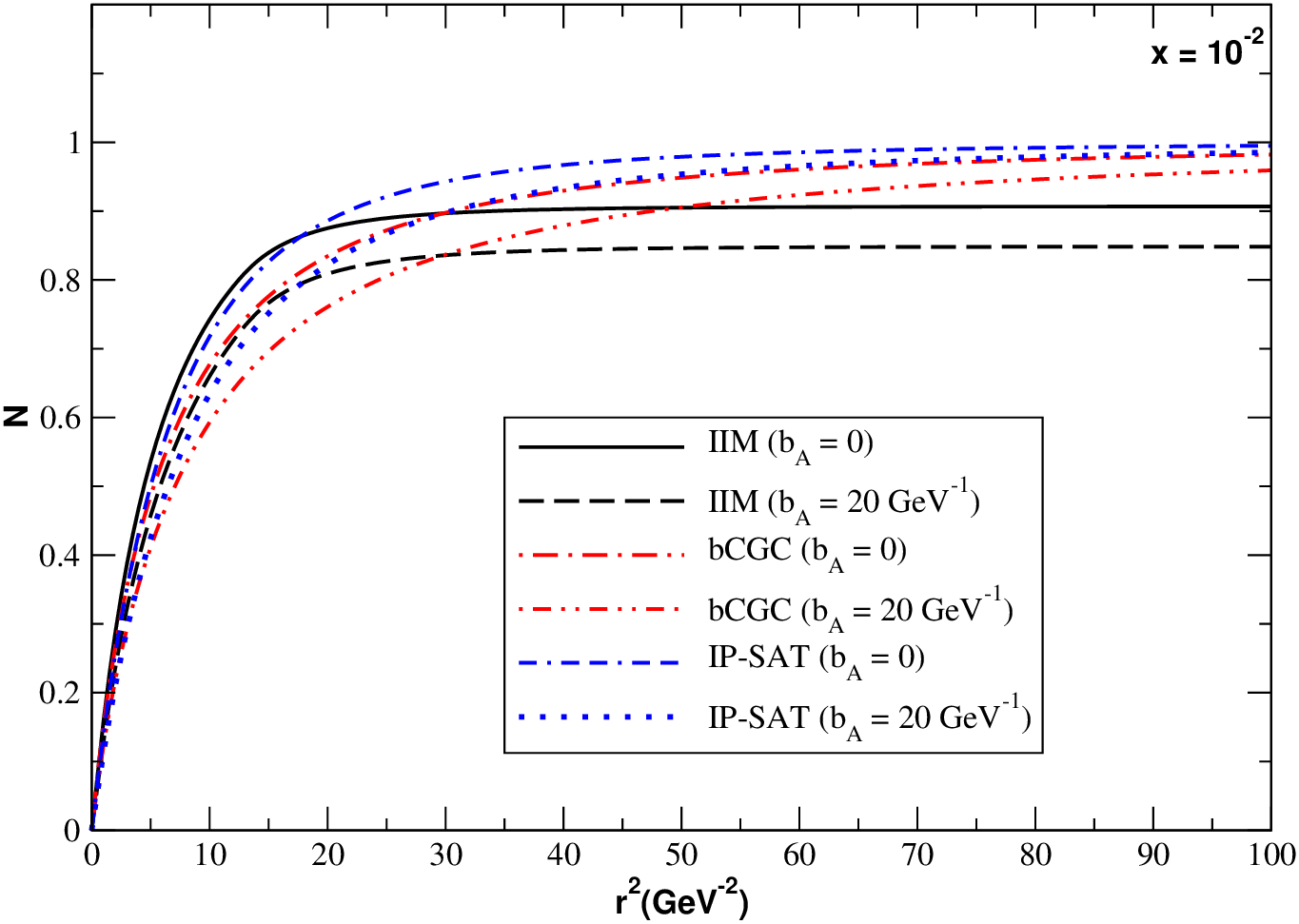}
}
{
\includegraphics[height=6cm]{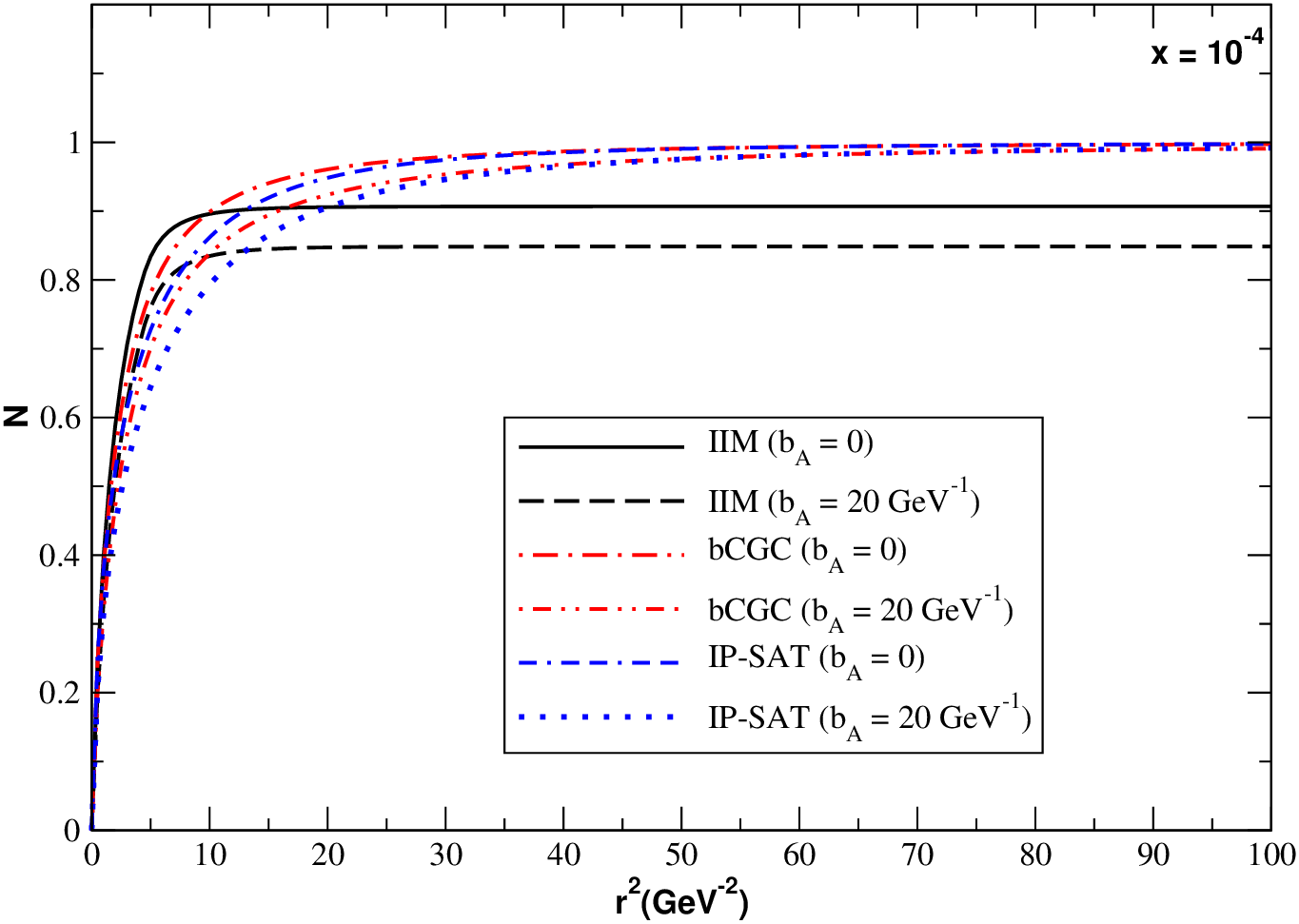}
}}
\end{tabular}
\caption{Dependence of dipole - nucleus scattering amplitude ${\cal N}_{A}$ as a function of $r^2$ for fixed values of $b$, $A = Pb$ and different values of $x$: $x=10^{-2}$ (left panel) and  $x=10^{-4}$ (right panel).}
\label{NA}
\end{figure}

How to treat the dipole - nucleus interaction  is still an open question 
due to the complexity of  the impact parameter dependence. In principle, it is possible to adapt the phenomenological models described above to the nuclear 
case (See e.g. Refs. \cite{armesto,simone,raju}) or to consider the numerical solution of the BK equation. In what follows, we will assume the model proposed in Ref. 
\cite{armesto}, which  includes the  impact parameter dependence and describes  
the existing  experimental data on the nuclear structure function \cite{erike}. 
In this model the  dipole-nucleus scattering amplitude is given by
\begin{eqnarray}
 {\cal N}_{A} = 1 - \exp \left[
-\frac{1}{2} \sigma_{dip}(x,r^{2}) \, T_{A}(\textbf{\textit{b}}_{A})
 \right] ,
 \label{Na_Glauber}
\end{eqnarray}
where
\begin{eqnarray}
 \sigma_{dip}(x,r^{2}) = 2 \int d^{2} \textbf{\textit{b}}_{p} \,\, {\cal N}_{p} 
(x,\textbf{\textit{r}},\textbf{\textit{b}}_{p}) ,
\end{eqnarray}
and $T_{A}(\textbf{\textit{b}}_{A})$ is the nuclear thickness, 
 which is obtained from a 3-parameter Fermi distribution for the nuclear
density normalized to $A$.
The above equation
sums up all the 
multiple elastic rescattering diagrams of the $q \overline{q}$ pair
and is justified for large coherence length, where the transverse separation $\rr$ of 
partons in the multiparton Fock state of the photon becomes a conserved quantity, {\it i.e.} the size of the pair $\rr$ becomes eigenvalue
of the scattering matrix.  In what follows we will compute  $\mathcal{N}_A$ considering 
the different models for the dipole - proton scattering amplitude discussed before.
In Fig. \ref{NA} we compare the predictions for the nuclear scattering amplitude considering the IIM, bCGC and IP-SAT models as input and different values of the impact parameter $b_{A}$. As expected, $\mathcal{N}_A$ saturates faster for central collisions than for large impact parameters. Moreover, the differences 
between the predictions are reduced in comparison to the proton case.
 This is directly associated with 
the model for ${\cal N}_{A}$, given by  Eq.(\ref{Na_Glauber}), which is the same 
in all three  cases. The future  experimental data on vector meson photoproduction in $PbPb$ collisions will be useful to test this model of the dipole - nucleus interaction.

Before presenting our results in the next section, some additional comments are in order. As the IIM model assumes the factorization of the impact parameter dependence, it implies that $\sigma_{dip}$ is given by
\begin{eqnarray}
  \sigma_{dip}(x,r^{2}) = 2 \int d^{2} \textbf{\textit{b}}_{p} \,\, {\cal N}_{p}(x,\mbox{\textbf{\textit{r}}}) \, S(
\mbox{\textbf{\textit{b}}}_{p}) = \sigma_{0} \,\,{\cal N}_p(x,\rr) ,
\end{eqnarray}
where $\sigma_{0}$ is a free parameter of the model constrained by the HERA data. In  order to estimate the vector meson cross section in $\gamma p $ interactions using this model we will assume an exponential Ansatz for the $t$ -- dependence of $\frac{d\sigma}{dt}$, which describes the typical behavior of a diffractive 
event. As a consequence, for the IIM model, we will have that
\begin{eqnarray}
 \sigma(\gamma h \rightarrow V h) = \left. \frac{1}{B_{V}} \frac{d\sigma}{dt} 
\right|_{t=0} ,
\label{cs_gammap_slope}
\end{eqnarray}
where $B_{V}$ is the slope parameter for the meson $V$. In our calculations, 
we have taken the slope parameters from Ref. \cite{glauber1}.
Finally, as in Refs. \cite{bruno1,bruno3} we have considered the real part 
of the amplitude and the skewedness correction in the $\gamma p$ case. According to   
Ref. \cite{Lappi}, the skewedness corrections are better justified at high energies.  
In the UPC nuclear case these corrections can be very large, probably because of the 
wide range of reached rapidities, which includes photon-target collisions at 
low energies. 
In \cite{Lappi} the authors argue that this can make  their cross sections  bigger than 
the measured ones for $J/\psi$ production in $Pb Pb$ collision. In the lack of a better  
understanding of the skewedness corrections in  nuclear collisions we will not include 
them in our study of $\gamma Pb$  scattering.

 \section{Results}
 \label{res}
 
One of the main observables that 
can be directly measured at the LHC and can be used to study  vector meson  
photoproduction is the rapidity distribution. Using the expression 
 \begin{eqnarray}
 \omega = \frac{M_{V}}{2} \exp \left( Y \right),
\end{eqnarray}
which relates the photon energy $\omega$,  the mass ($M_{V}$) and the  
rapidity ($Y$) of the produced meson in a UPC, and performing a change of  
variables in Eq. (\ref{epa}), we can show that  
\begin{eqnarray}
\frac{d\sigma \,\left[h_1 + h_2 \rightarrow   h_1 \otimes V \otimes h_2\right]}{dY} = \left[n_{h_{1}} (\omega) \,\sigma_{\gamma h_2 \rightarrow V \otimes h_2}\left(\omega \right)\right]_{\omega_L} + \left[n_{h_{2}} (\omega)\,\sigma_{\gamma h_1 \rightarrow V \otimes h_1}\left(\omega \right)\right]_{\omega_R}\,
\label{dsigdy}
\end{eqnarray}
where $\omega_L \, (\propto e^{+Y})$ and $\omega_R \, (\propto e^{-Y})$ denote photons from the $h_1$ and $h_2$ hadrons, respectively.
In what follows, we will present our results for the rapidity distribution of  
vector mesons produced  in UPC's. In our predictions we did not  
consider the corrections associated to soft interactions which would destroy the rapidity gaps \cite{Schafer,Martin} and, in the 
nuclear case, we did not include possible gluon shadowing corrections \cite{Guzey,glauber1}. The treatment of both corrections is still is a theme of debate. 
In principle, these two  corrections 
would lead to a reduction of the cross sections.


\begin{figure}[!t]
\begin{tabular}{ccc}
\centerline{
{
\includegraphics[height=4.5cm]{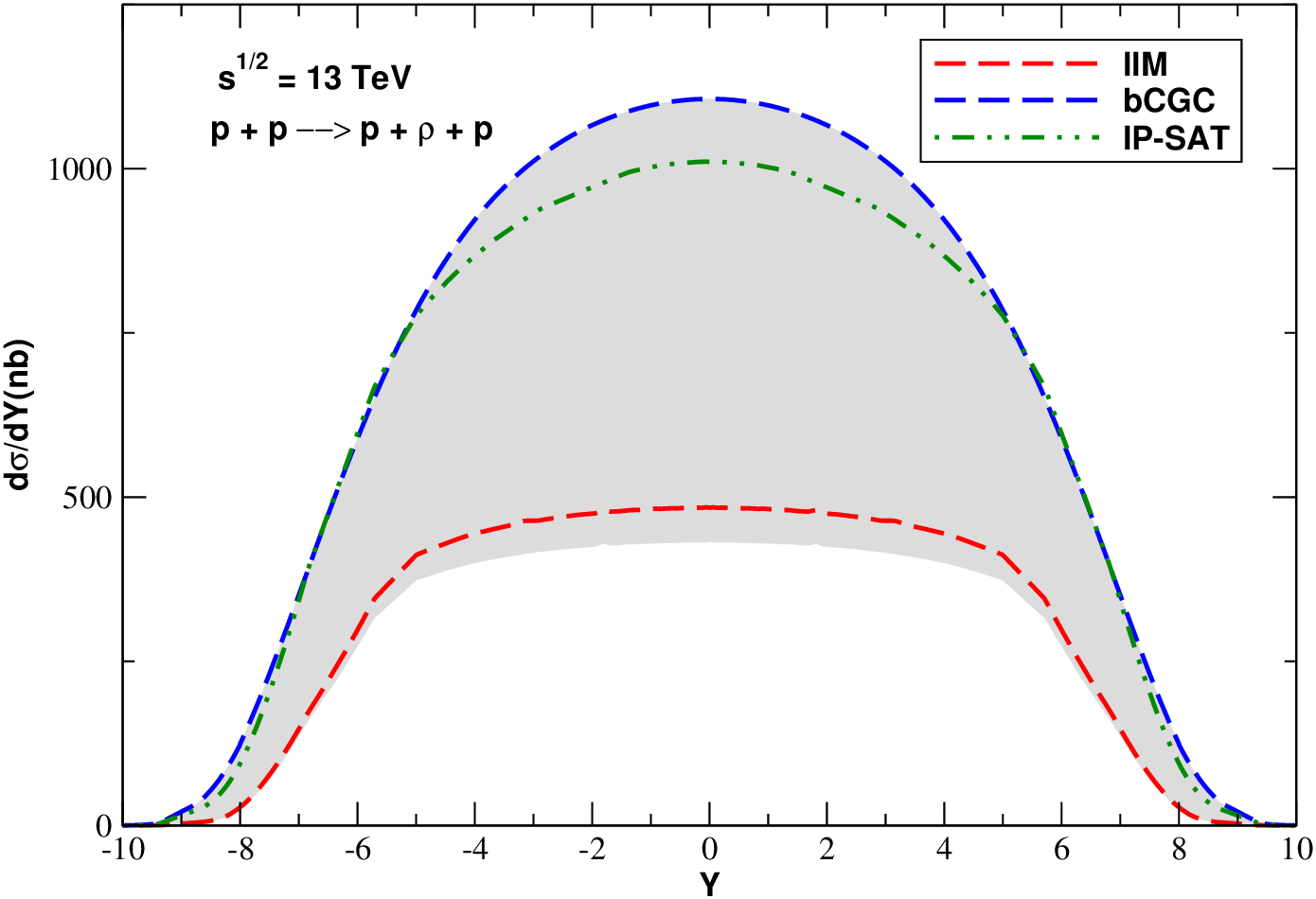}
}
{
\includegraphics[height=4.5cm]{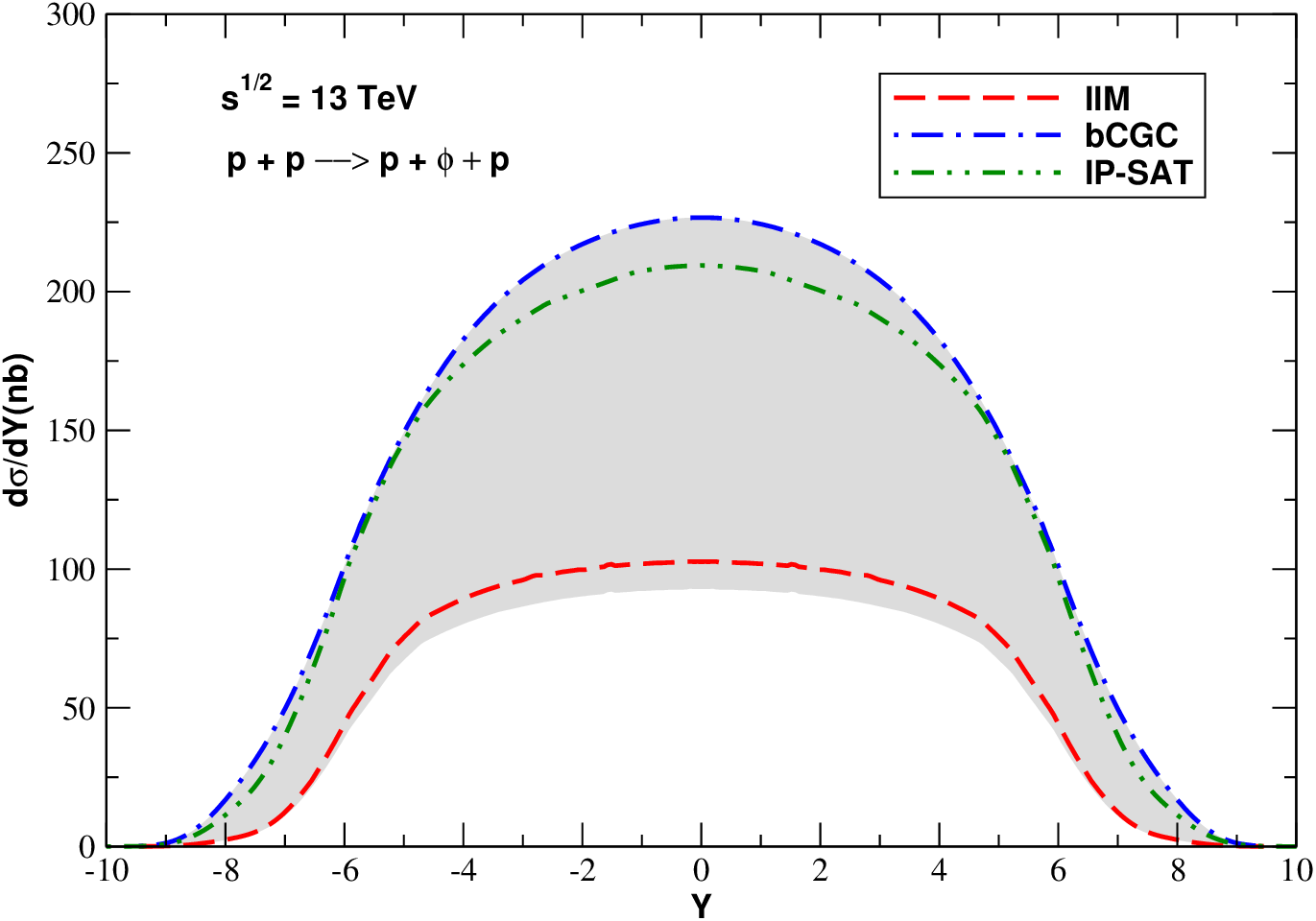}
}
}
{
}
\end{tabular}  \\
\begin{tabular}{ccc}
\centerline{
{
\includegraphics[height=4.5cm]{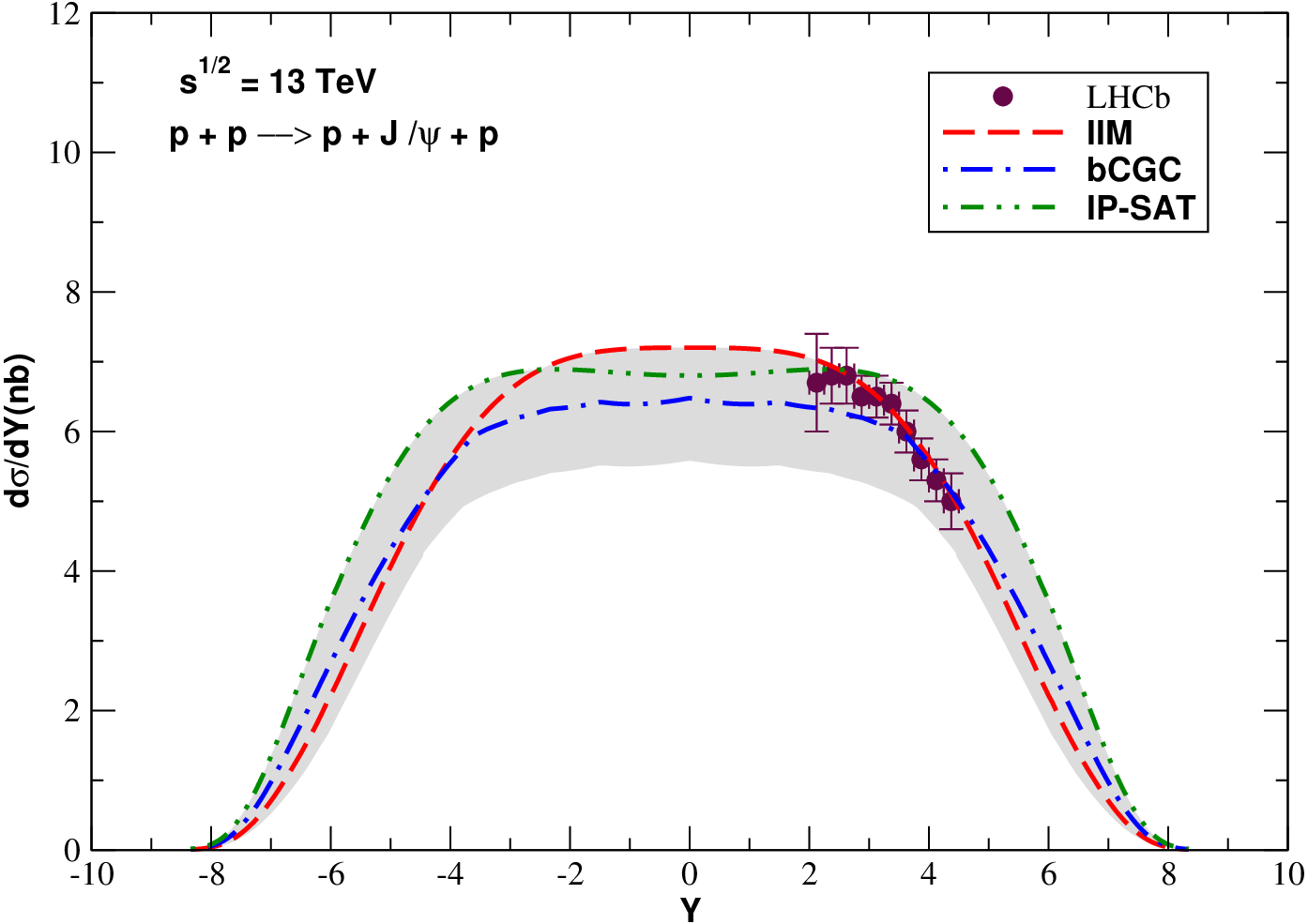}
}
{
\includegraphics[height=4.5cm]{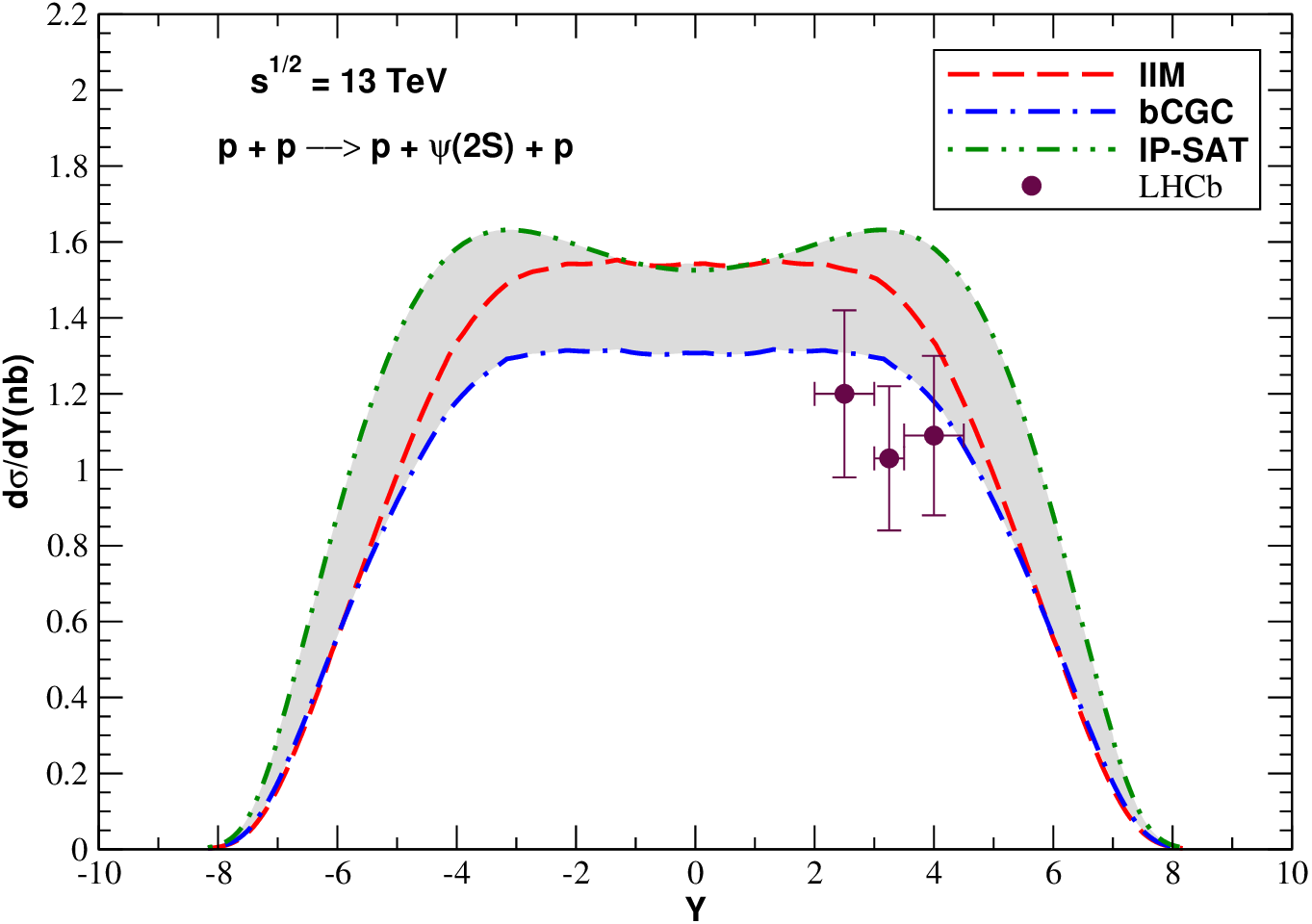}
}
{
\includegraphics[height=4.5cm]{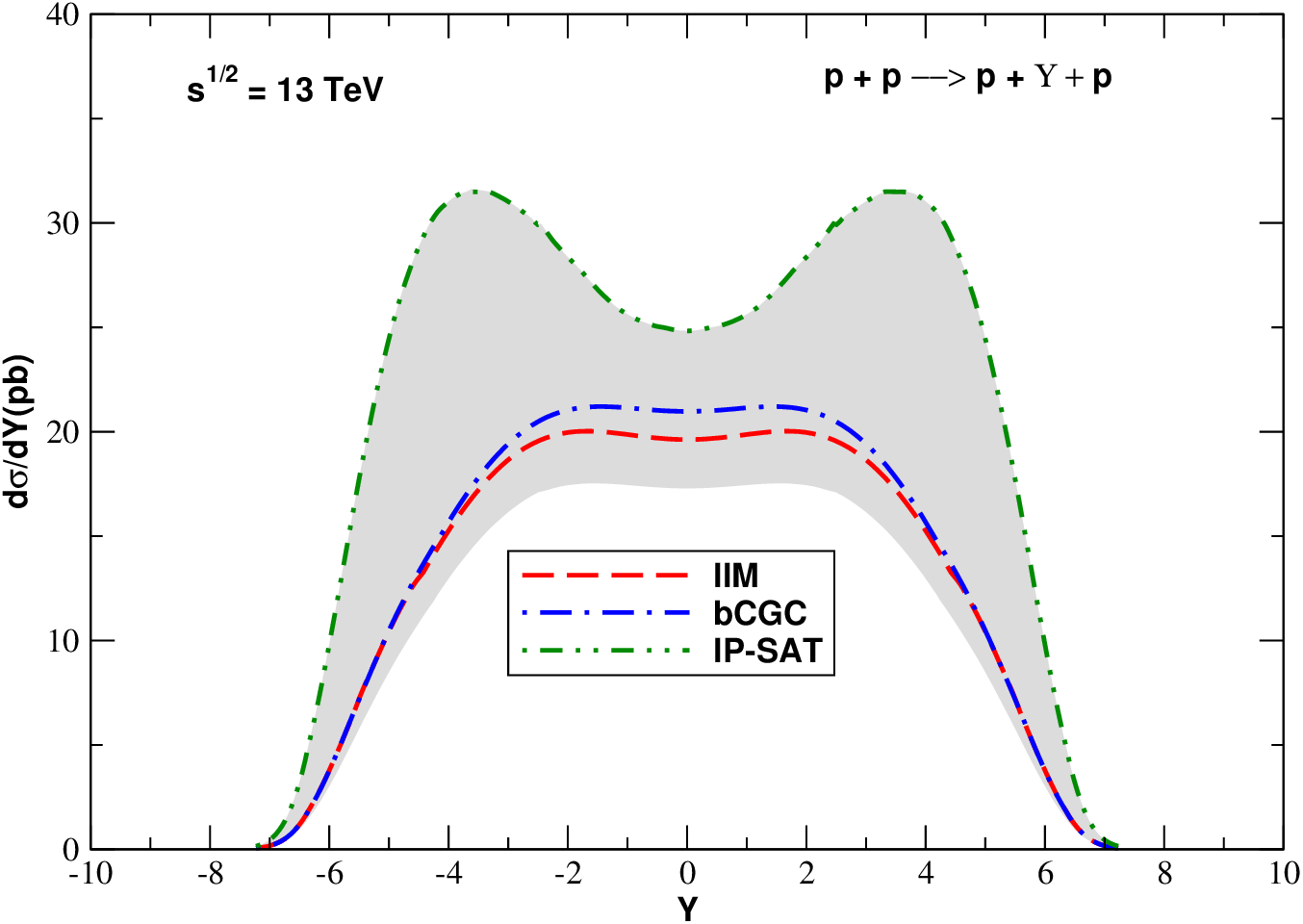}
}}
\end{tabular}
\caption{Rapidity distributions for  exclusive photoproduction of $\rho$, $\phi$, $J/\psi$, $\psi(2S)$ and $\Upsilon$  in $pp$ collisions at $\sqrt{s} = 13$ TeV. Data from LHCb Collaboration \cite{lhcb,lhcb2,lhcb3}. }
\label{pp}
\end{figure}

In order to estimate the dependence of our predictions on the model used to describe the dipole -- target scattering amplitude, let us focus initially on 
the rapidity distributions for the photoproduction of $\rho$, $\phi$, $J/\psi$, $\psi(2S)$ and $\Upsilon$ in $pp$ collisions at $\sqrt{s} = 13$ TeV considering the three models for ${\cal N}_p$ discussed in the previous section and the Boosted Gaussian model for $\psi_{V}$. The different lines present in the Fig. \ref{pp} represent the predictions of the IIM, bCGC and IP-SAT models. The results obtained  with the updated IP-SAT model are shown here for the  first time.
 As discussed before, the overlap functions for the $\rho$ and $\phi$ mesons are dominated by  larger dipole sizes than for heavier mesons ($J/\Psi$, $\Psi(2S)$ and $\Upsilon$). Therefore, the light and heavy meson states probe ${\cal N}_h$ at different values of $\rr$. We observe that for $\rho$ production,  the IP-SAT and the bCGC models give very similar predictions while the IIM prediction is smaller  by a factor 2. This result is directly associated to the behavior of ${\cal N}_p$ at large dipoles presented in Fig. \ref{Np}.   On the other hand, in the $\Upsilon$ production we are probing smaller dipoles, where the difference between the IIM, bCGC and IP-SAT models is associated to the distinct description of the linear regime. In this case, the IIM and bCGC predictions are similar and the IP-SAT predicts larger value of the rapidity distribution at central rapidities. The different modelling of the linear and nonlinear regimes in the distinct models of ${\cal N}_p$, as well as of the transition between these regimes, has direct impact on the predictions for the different mesons, as can be observed in Fig. \ref{pp}. While the IIM prediction is a lower bound for the $\rho$ meson production at midrapidity, it is an upper bound for the $J/\Psi$ one and becomes a lower bound for the $\Upsilon$ production. Therefore, a global analysis of different final states is an important probe of the treatment of QCD dynamics in the linear and nonlinear regimes. 
 
In order to estimate the  theoretical uncertainty in the color dipole predictions for  exclusive vector meson photoproduction in $pp$ collisions, we also include in Fig. \ref{pp} a band which appears when  we combine the predictions from 
the three different models of ${\cal N}_p$ with the two models of  $\psi_{V}$.  
In the  $J/\psi$ and $\psi (2S)$ cases 
we find a good agreement between the predictions and the data from LHCb Collaboration 
\cite{lhcb,lhcb2,lhcb3}.  Moreover, the uncertainty in the 
predictions is bigger for light vector mesons at $Y=0$, reaching a factor 3. 
Additionally, a large uncertainty is present in the predictions for the $\Upsilon$ production at large - $Y$.
  
 \begin{figure}[!t]
\begin{tabular}{ccc}
\centerline{
{
\includegraphics[height=4.5cm]{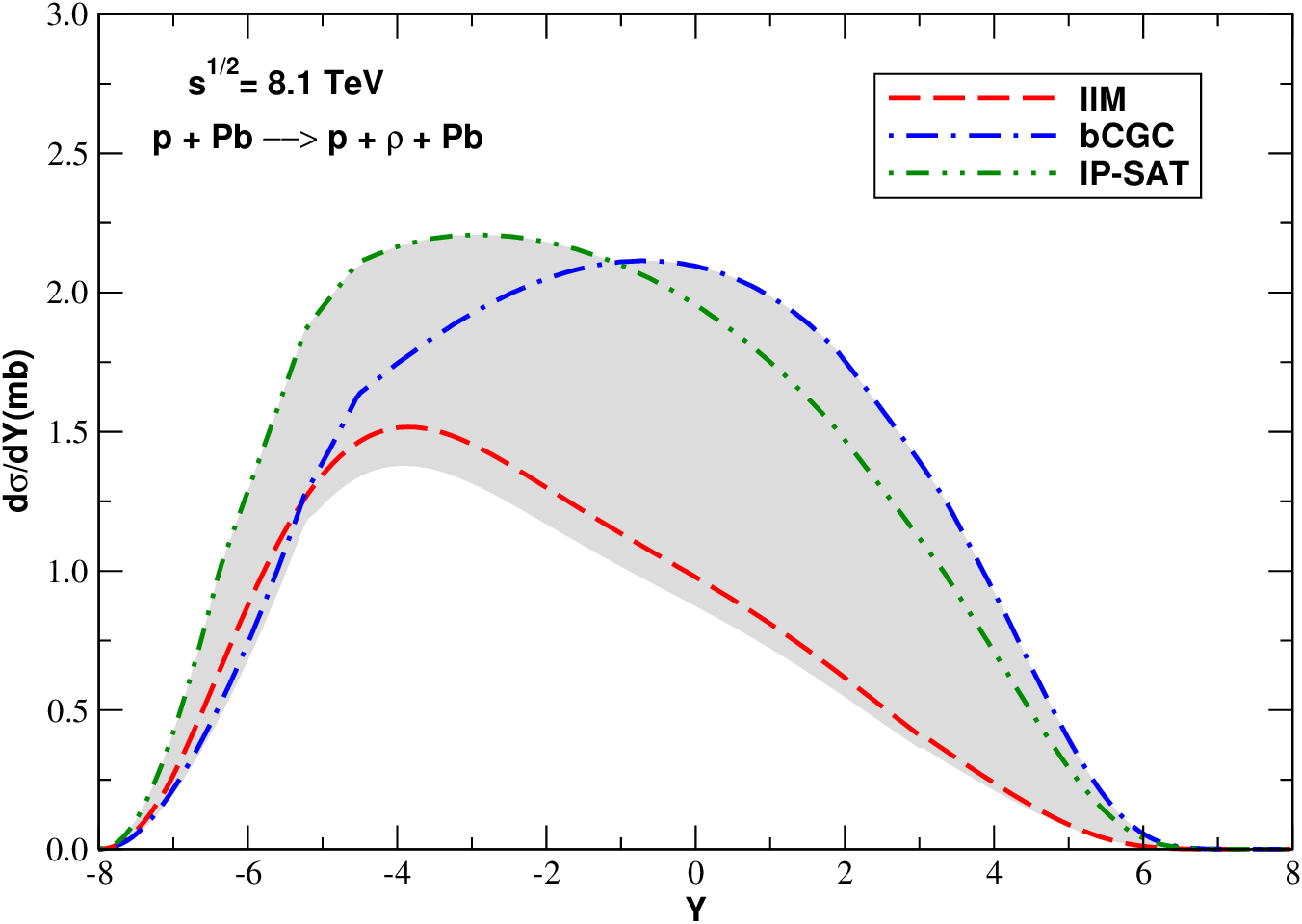}
}
{
\includegraphics[height=4.5cm]{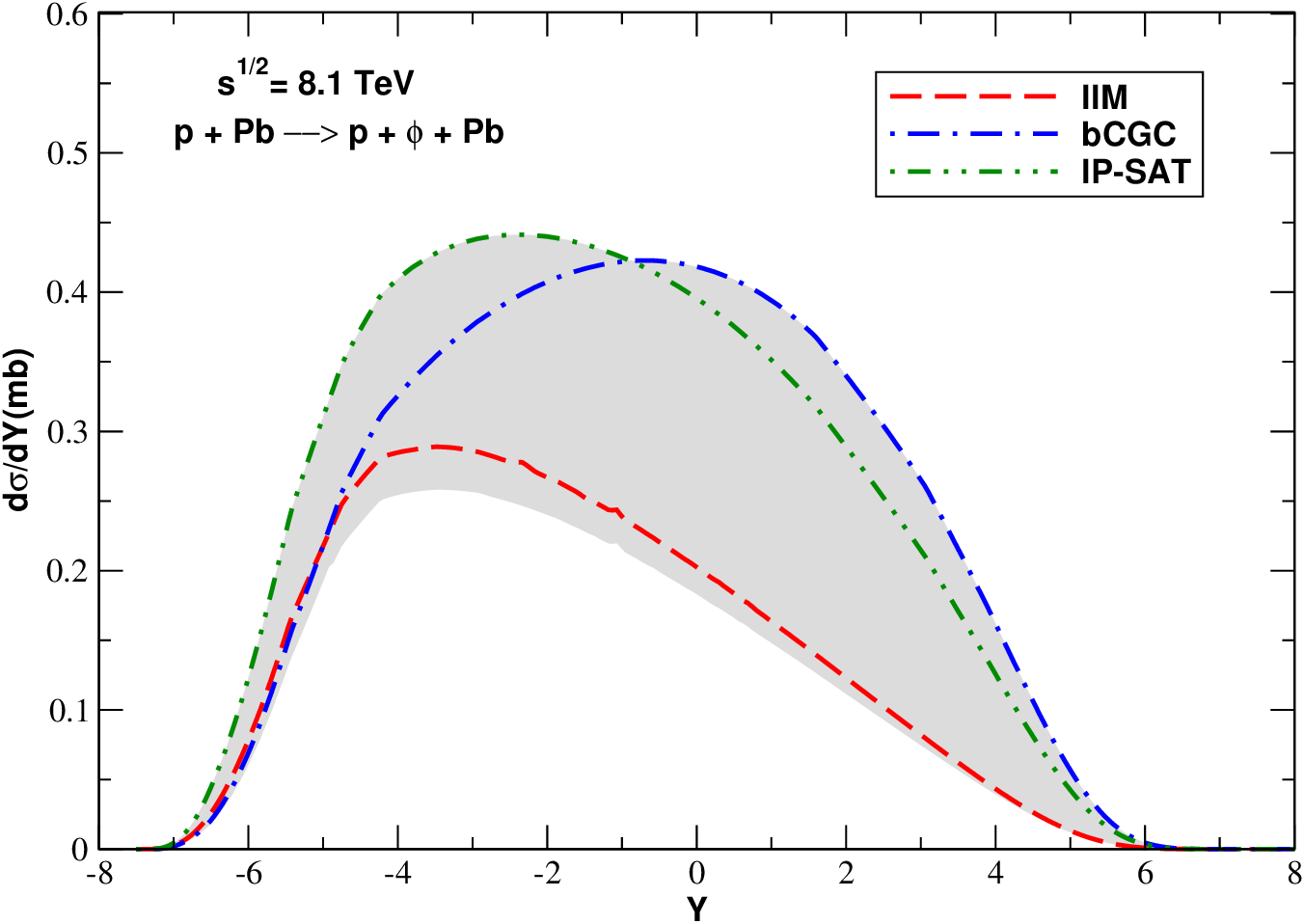}
}
}
\end{tabular}  \\
\begin{tabular}{ccc}
\centerline{
{
\includegraphics[height=4.5cm]{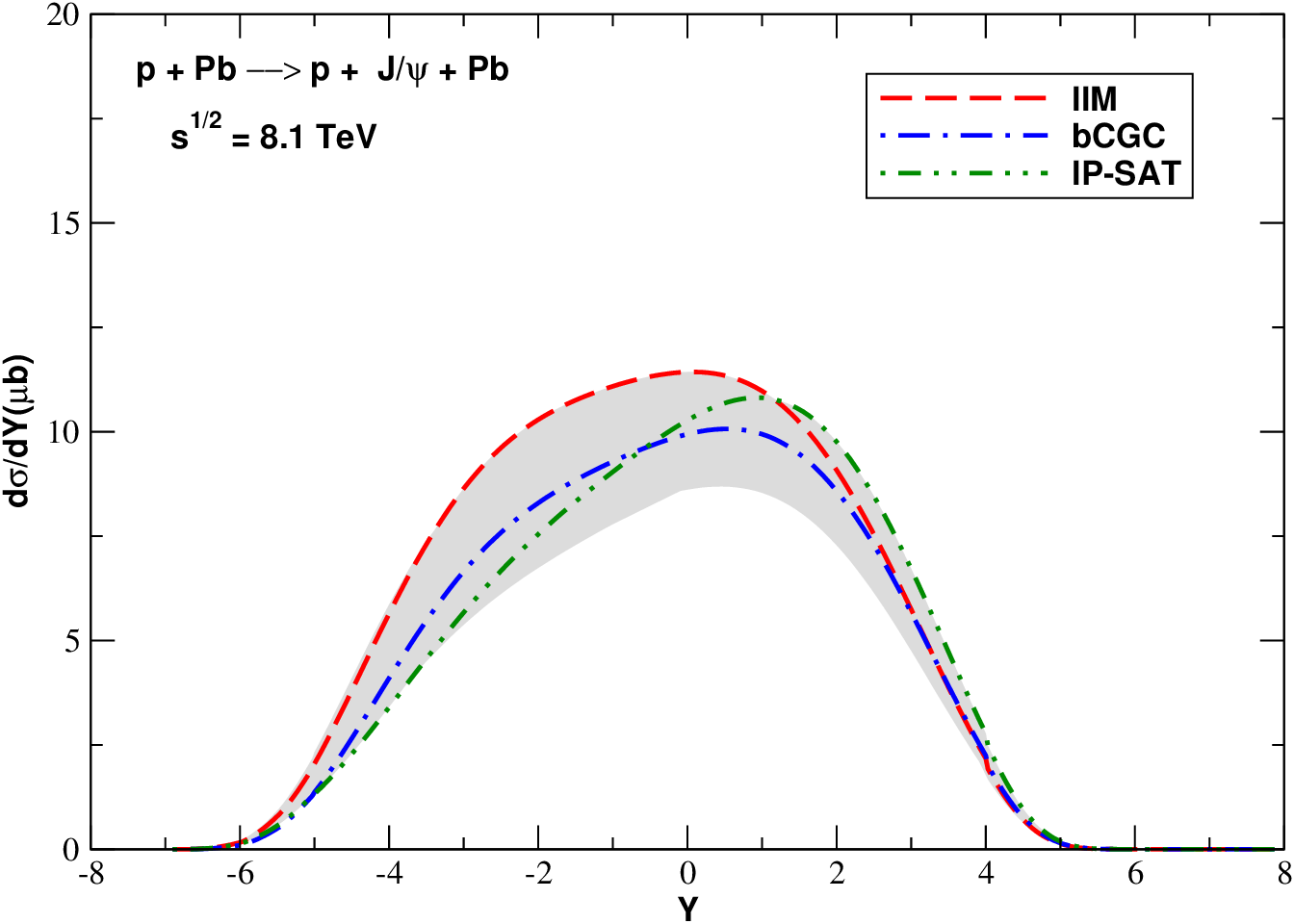}
}
{
\includegraphics[height=4.5cm]{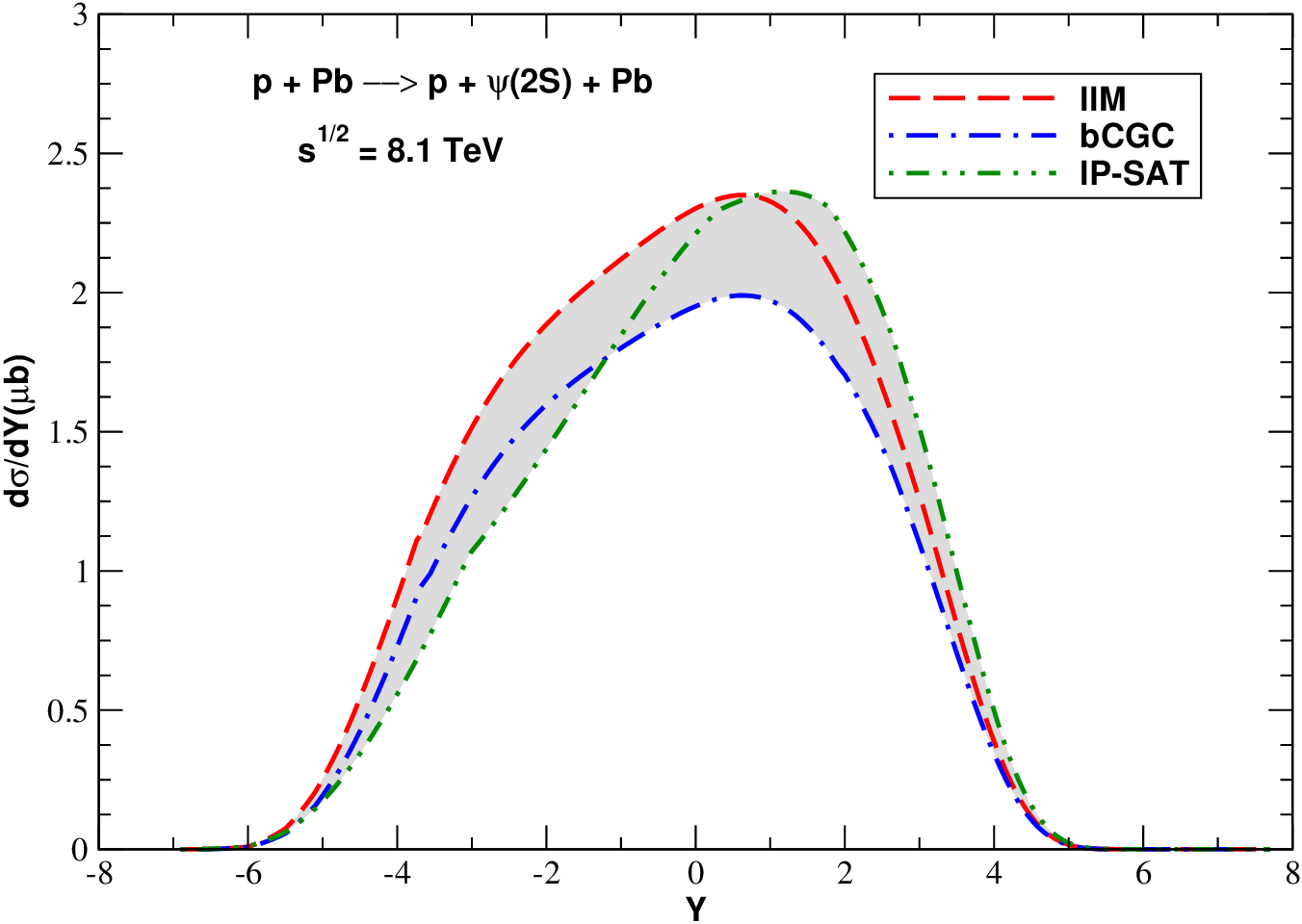}
}
{
\includegraphics[height=4.5cm]{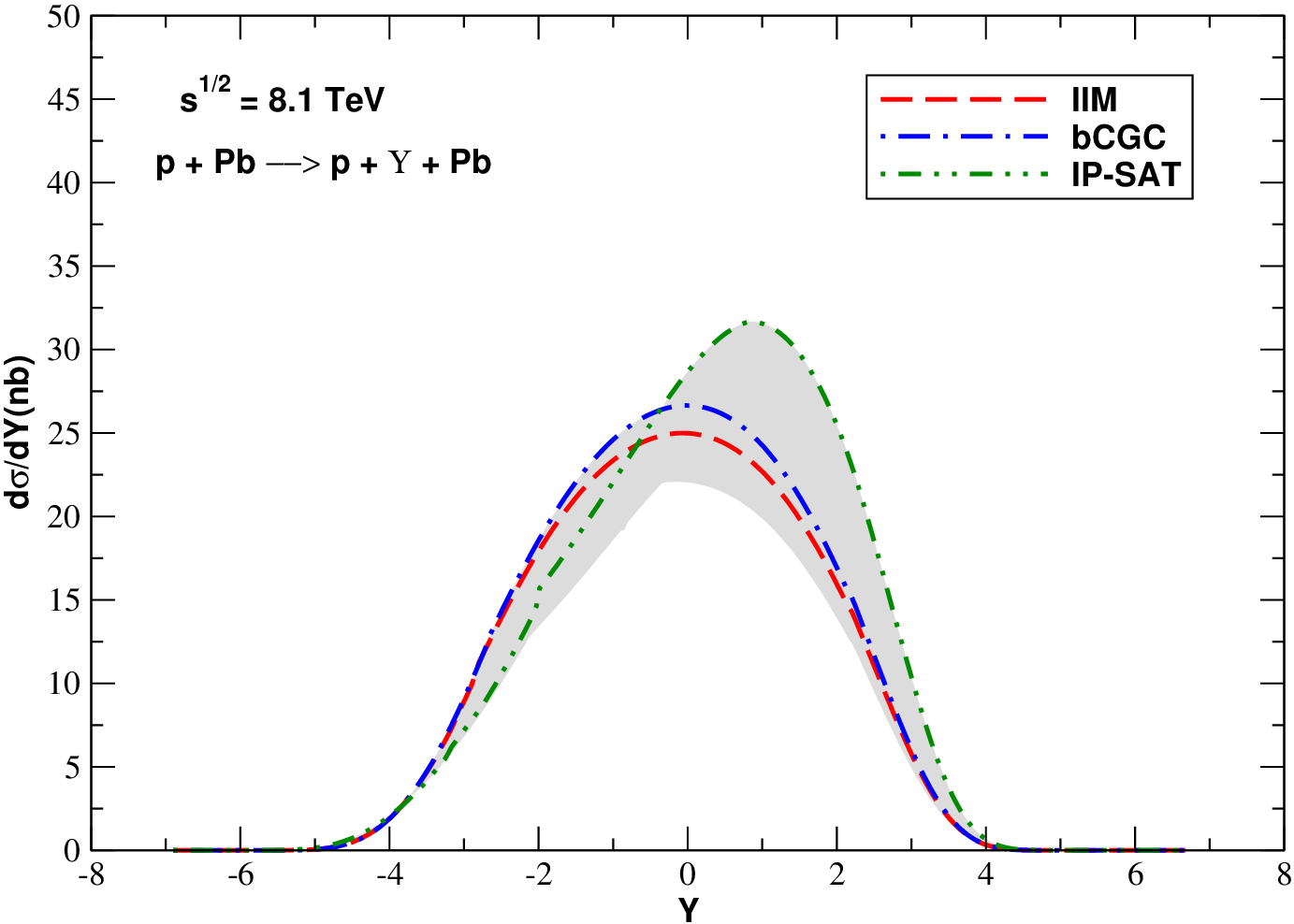}
}}
\end{tabular}
\caption{Rapidity distributions for the exclusive photoproduction of $\rho$, $\phi$, $J/\psi$, $\psi(2S)$ and $\Upsilon$ in $pPb$ collisions at $\sqrt{s} = 8.1$ TeV. }
\label{pA}
\end{figure}

Let us now consider $pPb$ collisions. In this case   
the  $\gamma p$ interactions are dominant because the equivalent photon 
spectrum of the nucleus is enhanced by a factor $Z^{2}$ in comparison to 
the proton one. Consequently, the rapidity distribution is asymmetric with respect to $Y = 0$. One advantage of the study of $pPb$ collisions is that the analysis of the rapidity distribution for a given value of $Y$ gives direct access to the value of $x$ probed in the scattering amplitude, in contrast to symmetric collisions, which receive contributions of the QCD dynamics at small and large values of $x$. As the behavior of 
${\cal N}_p$ at large - $x$ is not under theoretical control, it has direct impact on the color dipole predictions for symmetric collisions. An asymmetric distribution is observed in Fig. \ref{pA}, where we present our predictions for the
exclusive photoproduction of $\rho$, $\phi$, $J/\psi$, $\psi(2S)$ and $\Upsilon$ in $pPb$ collisions at $\sqrt{s} = 8.1$ TeV. 
As in the $pp$ case, the IIM, bCGC and IP-SAT predictions differ significantly in the production of light vector mesons, which implies a large theoretical uncertainty. For heavy mesons, the uncertainty is smaller, but still significant for  $\Upsilon$ production at forward rapidities. Finally, it is important to emphasize that the position of the maximum of the distribution is model and vector meson dependent. 
This fact can be used to test details of the QCD dynamics in a future global 
analysis of  exclusive vector meson photoproduction in $pPb$ collisions.

 \begin{figure}[!t]
\begin{tabular}{ccc}
\centerline{
{
\includegraphics[height=4.5cm]{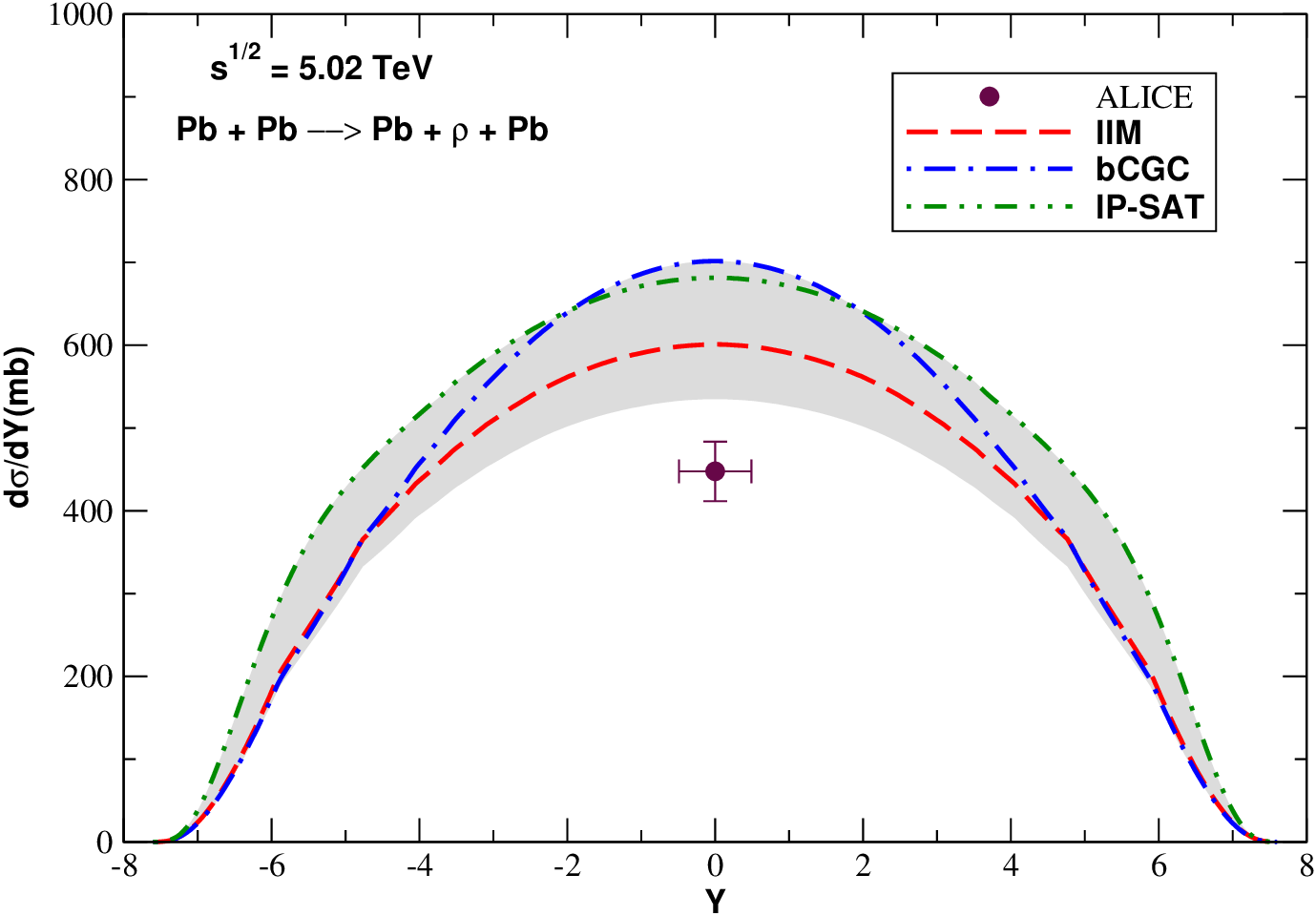}
}
{
\includegraphics[height=4.5cm]{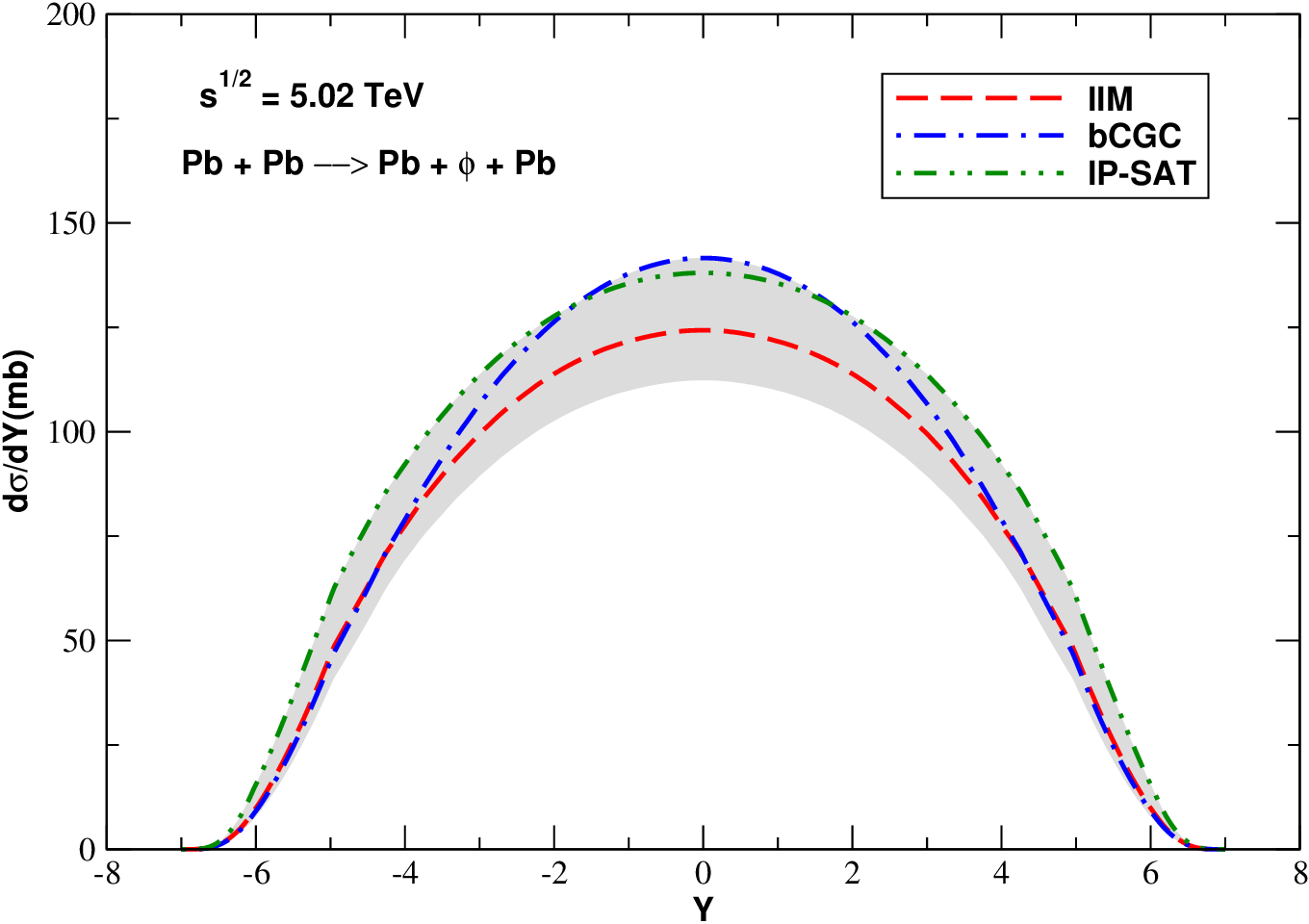}
}
}
\end{tabular}  \\
\begin{tabular}{ccc}
\centerline{
{
\includegraphics[height=4.5cm]{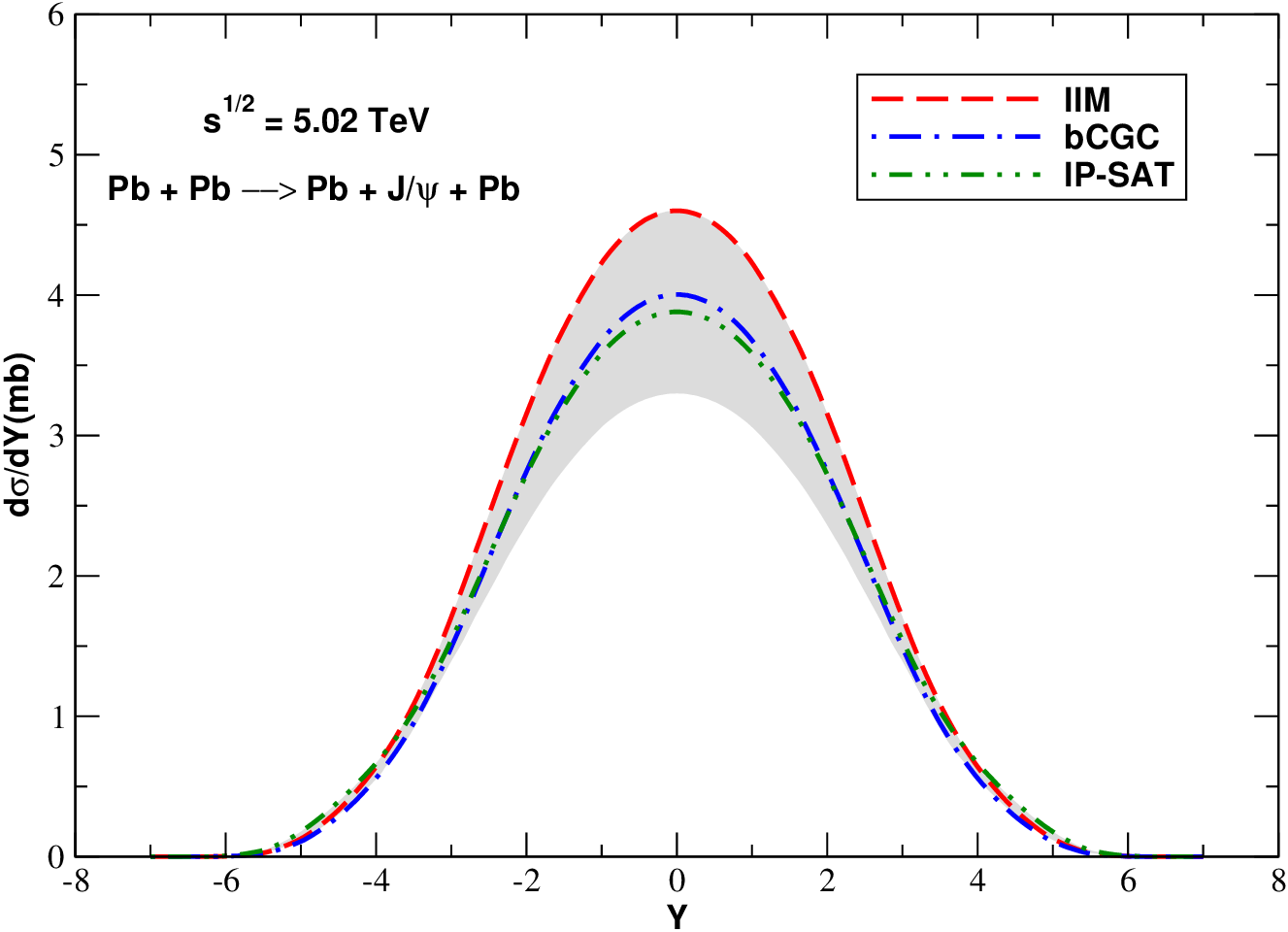}
}
{
\includegraphics[height=4.5cm]{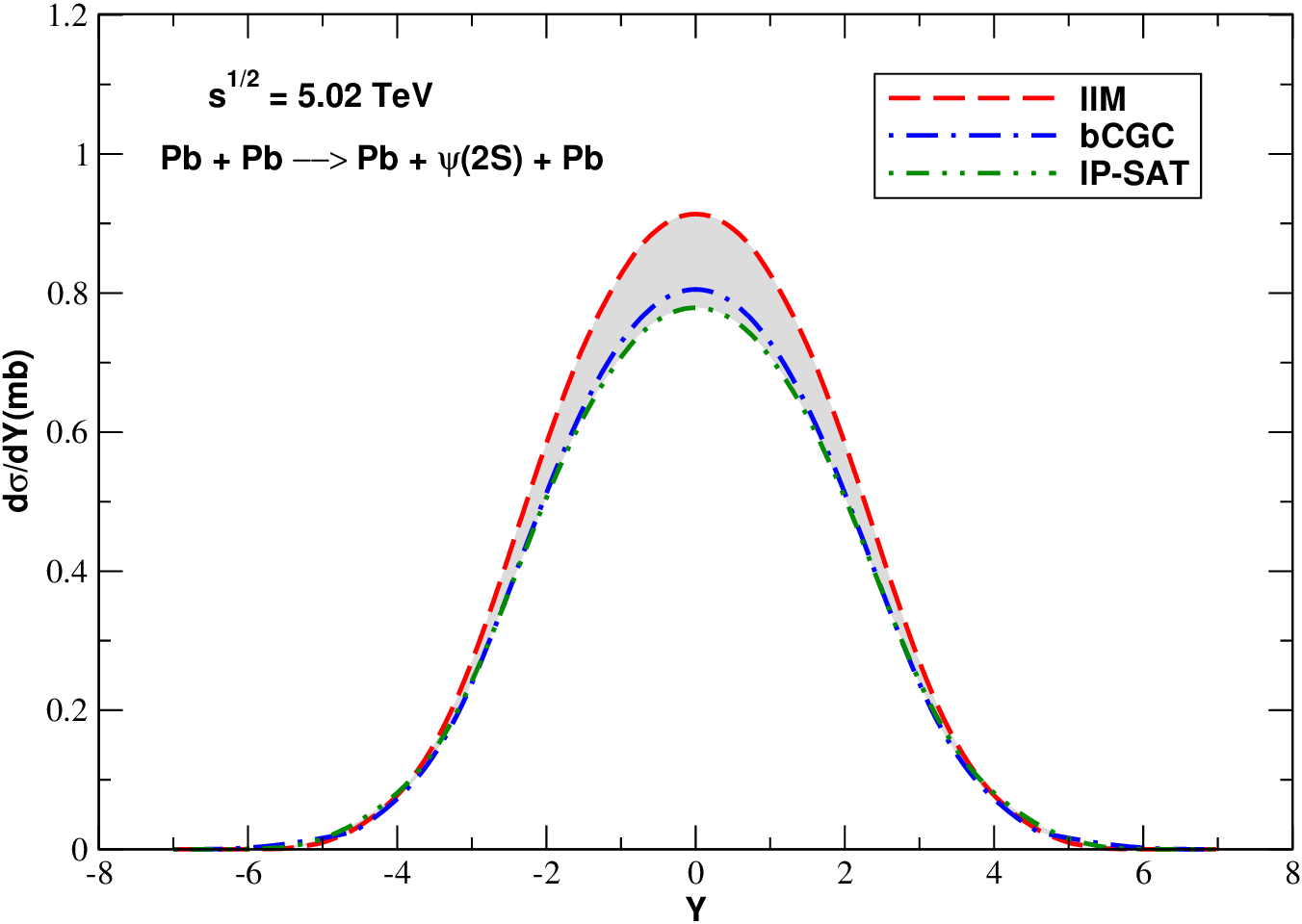}
}
{
\includegraphics[height=4.5cm]{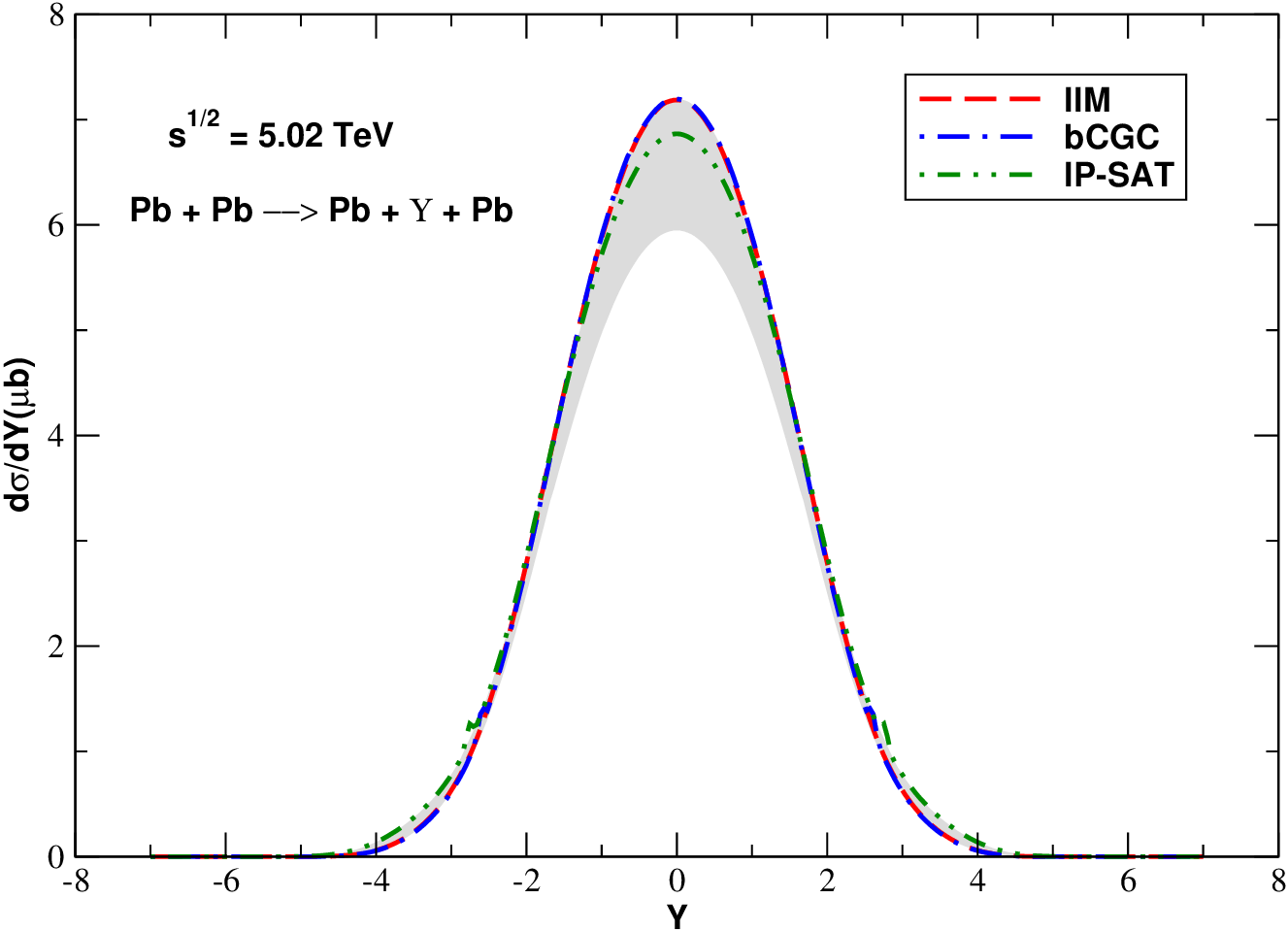}
}}
\end{tabular}
\caption{Rapidity distributions for the exclusive photoproduction of $\rho$, $\phi$, $J/\psi$, $\psi(2S)$ and $\Upsilon$ in $PbPb$ collisions at $\sqrt{s} = 5.02$ TeV. Data are from ALICE Collaboration  \cite{alice,alice2}.  }
\label{AA}
\end{figure}

In Fig. \ref{AA} we present our predictions for the exclusive photoproduction of $\rho$, $\phi$, $J/\psi$, $\psi(2S)$ and $\Upsilon$ in $PbPb$ collisions at $\sqrt{s} = 5.02$ TeV. In this case the difference between the predictions is smaller in comparison to the $pp$ case, as expected from Fig. \ref{NA}.  In particular, 
 the   bCGC and IP-SAT predictions are similar at central  rapidities. 
Recently the ALICE Collaboration has released new data on  
$\rho$ photoproduction at central rapidity \cite{alice,alice2}, which is presented in Fig. \ref{AA}.  As can be observed, the color dipole predictions overestimate 
the data. This result is an indication that other effects, not included 
in our analysis, should be included at least for this final state. Some possibilities are the inclusion of  
shadowing corrections \cite{glauber1} or  absorption corrections 
\cite{Martin}. Another possible conclusion is that the treatment of 
the dipole -- nucleus interaction, described here by the model presented 
in Eq. (\ref{Na_Glauber}), should be improved. Surely, new data on other 
mesons species will clarify this point. It may be possible to see if this 
failure is  only related to  $\rho$ production or if it is an 
indication of a limitation of the existing color dipole descriptions.

Finally, for completeness, we present 
in Table \ref{sec_tot_tab} the lower and upper bounds of our predictions for the total cross sections, obtained considering the three models for the dipole - hadron scattering amplitude and the two models for the vector meson wave functions.  As expected from the analysis of the rapidity distributions, we observe that $\rho$ 
production suffers from the largest uncertainties. Additionally, the cross sections 
are larger for $PbPb$ collisions and decrease with the mass of the vector meson.

\begin{table}[!tb]
\centering

\begin{small} 
  
\setlength{\tabcolsep}{3pt} 


\begin{tabular}{|c|c|c|c|c|c|c|c|c|c|c|c|c|c|c|c|c|c|c|c|c|c|c|c|c|}\hline
\raisebox{1.5ex}{}         & $\rho$            & $\phi$           & $J/\psi$           & $\psi(2S)$         & $\Upsilon$ \\ \hline

pp ($\sqrt{s} = 13$ TeV)   &5.37 - 13.03 $\mu$b&1.03 - 2.52 $\mu$b&58.74 - 82.90 nb    &14.81 - 19.31 nb    &0.17 - 0.33 nb 	 \\ \hline
pPb ($\sqrt{s} = 8.1$ TeV)  &   9.78 - 20.94 mb &   1.79 - 3.79 mb &55.78 - 80.13 $\mu$b&12.12 - 15.61 $\mu$b&0.10 - 0.15  $\mu$b  \\ \hline
PbPb ($\sqrt{s} = 5.02$ TeV) &   5.26 - 7.04 b   &   0.94 - 1.23 b  &18.24 - 24.47 mb    &3.85 - 4.47 mb      &22.20 - 26.48 $\mu$b  \\ \hline

\end{tabular} 
\end{small}
\caption{Lower and upper bounds for the total cross sections of the exclusive vector meson photoproduction in $pp/pPb/PbPb$ collisions at the Run 2 LHC energies.}
\label{sec_tot_tab}
\end{table}

\section{Conclusions}
\label{conc}
Recent studies show that vector meson exclusive photoproduction has the  
potential to probe the QCD dynamics at high energies. This will be even 
more so  in the forthcoming Run 2, when the total luminosity will be much 
higher than in Run 1.  This larger data sample will allow the study of a  
larger set of different final states and  a better discrimination between 
alternative descriptions.

In this paper we have presented a comprehensive study of the light and heavy  
vector meson photoproduction in $pp/pPb/PbPb$ collisions at Run 2 LHC energies 
using the  color dipole formalism.  We have used the updated versions of different 
models of the dipole scattering amplitude, which take into account   the 
nonlinear effects of the QCD dynamics (which are expected 
to become visible at the currently available energies). Moreover, we have used two 
different vector meson wave functions which, together with the models for the amplitude, describe the  HERA data on vector meson production. As the LHC probes a larger range of $\gamma h$ center of mass energies, the analysis of   
vector meson photoproduction in this collider can be useful to  
discriminate between these distinct descriptions of the QCD dynamics 
and vector meson formation.
As the free parameters present in the color dipole formalism have been    
constrained by the HERA data, the predictions for LHC energies are parameter  
free. In our study we have presented predictions for the exclusive  
photoproduction of $\rho$, $\phi$, $J/\psi$, $\psi(2S)$ and $\Upsilon$ 
in $pp/pPb/PbPb$ collisions using the IIM, bCGC and IP-SAT models  for 
the dipole - proton scattering amplitude and the Boosted Gaussian and  
Gaus - LC models for the vector meson wave function. Our results  
demonstrated that the light meson production probes larger dipole  
sizes and, consequently, the QCD dynamics deeper in the saturation 
regime. On the other hand, the heavy meson production probes the 
linear regime. These results indicated that a combined  
study of different final states will lead to a better                   
understanding  of  the transition from  linear to nonlinear dynamics. 
This study is very important to constraint the QCD dynamics, the vector meson wave function and the  
treatment of the nuclear interactions.
Our results are in  good agreement with the experimental data in the case 
of heavy vector meson production. The comparison with the recent experimental  
data on the $\rho $ production in $Pb Pb$ collisions indicated that the color 
dipole  description, with the current assumptions, overestimate the data. 
In principle, this can be  interpreted as an indication that a  more careful 
treatment of the dipole -- nucleus interaction and/or next - to - leading 
order corrections may be required and/or that shadowing effects and 
absorptive corrections should be incorporated to the formalism. 
Future experimental data may decide whether improvements of the color  
dipole description should also be included in the case of production of 
other vector mesons. 
Finally, it is important to emphasize that  the color dipole formalism has  
been recently extended to describe double vector meson photoproduction in 
hadronic collision \cite{bruno_double} and also to describe vector meson 
photoproduction associated to a leading neutron \cite{diego_leading}. As 
demonstrated in Refs. \cite{bruno_double,diego_leading}, these processes 
can in principle be studied in the LHC considering the Run 2 energies.  
As the basic ingredients of these calculations are the same used in the 
present paper, the analysis of these processes also can be useful to test 
the universality of the color dipole description.


\section*{Acknowledgements}
 This work was partially financed by the Brazilian funding agencies CAPES, CNPq,  
FAPESP (process number 12/50984-4), FAPERGS and INCT-FNA (process number 
464898/2014-5).




\begin{thebibliography}{99}

\bibitem{hdqcd} 
  F.~Gelis, E.~Iancu, J.~Jalilian-Marian and R.~Venugopalan,
    Ann.\ Rev.\ Nucl.\ Part.\ Sci.\  {\bf 60}, 463 (2010);
  H.~Weigert,  Prog.\ Part.\ Nucl.\ Phys.\  {\bf 55}, 461 (2005); J.~Jalilian-Marian and Y.~V.~Kovchegov, Prog.\ Part.\ Nucl.\ Phys.\  {\bf 56}, 104 (2006).


\bibitem{eics}
A.~Deshpande, R.~Milner, R.~Venugopalan and W.~Vogelsang,
  Ann.\ Rev.\ Nucl.\ Part.\ Sci.\  {\bf 55},  165 (2005);  D.~Boer, M.~Diehl, R.~Milner, R.~Venugopalan, W.~Vogelsang, D.~Kaplan, H.~Montgomery and S.~Vigdor {\it et al.},
  arXiv:1108.1713 [nucl-th];
  A.~Accardi, J.~L.~Albacete, M.~Anselmino, N.~Armesto, E.~C.~Aschenauer, A.~Bacchetta, D.~Boer and W.~Brooks {\it et al.},
 Eur.\ Phys.\ J.\ A {\bf 52}, no. 9, 268 (2016);
  E.~C.~Aschenauer {\it et al.},
  arXiv:1708.01527 [nucl-ex].


 \bibitem{upc}              G. Baur, K. Hencken, D. Trautmann, S. Sadovsky, Y. Kharlov, 
                            Phys. Rep. {\bf 364}, 359 (2002); 
                            V.~P.~Goncalves and M.~V.~T.~Machado,
                            Mod. Phys. Lett. A {\bf 19}, 2525  (2004); 
                            C.~A. Bertulani, S.~R.~Klein and J.~Nystrand, 
                            Ann. Rev. Nucl. Part. Sci. {\bf 55}, 271 (2005);
                            K.~Hencken {\it et al.}, 
                            Phys.\ Rept.\  {\bf 458}, 1 (2008);   J.~G.~Contreras and J.~D.~Tapia Takaki,
                            Int.\ J.\ Mod.\ Phys.\ A {\bf 30}, 1542012 (2015).


  
\bibitem{klein_prc}           S. R. Klein, J. Nystrand,  
                              Phys. Rev. C {\bf 60}, 014903 (1999).

\bibitem{gluon}               V.~P.~Goncalves and C.~A.~Bertulani, 
                              Phys.\ Rev.\ C {\bf 65}, 054905 (2002).

\bibitem{strikman}            L.~Frankfurt, M.~Strikman and M.~Zhalov,
                              Phys.\ Lett.\ B {\bf 540}, 220 (2002).


\bibitem{outros_klein}        S. R. Klein, J. Nystrand,   
                              Phys.\ Rev.\ Lett.\  {\bf 92}, 142003 (2004).


\bibitem{vicmag_mesons1}      V.~P.~Goncalves and M.~V.~T.~Machado,
                              Eur.\ Phys.\ J.\  C {\bf 40}, 519 (2005).
  
   
\bibitem{outros_vicmag_mesons}   V.~P.~Goncalves and M.~V.~T.~Machado, 
                                 Phys.\ Rev.\  C {\bf 73}, 044902 (2006); 
                                 Phys.\ Rev.\  D {\bf 77}, 014037 (2008); 
                                 Phys.\ Rev.\ C {\bf 80}, 054901 (2009).
  

\bibitem{outros_frankfurt}       L.~Frankfurt, M.~Strikman and M.~Zhalov,
                                 Phys.\ Lett.\ B {\bf 537}, 51 (2002); 
                                 Phys.\ Rev.\ C {\bf 67}, 034901 (2003);  
                                 L.~Frankfurt, V.~Guzey, M.~Strikman and M.~Zhalov, 
                                 JHEP {\bf 0308}, 043 (2003).


\bibitem{Schafer}         W.~Schafer and A.~Szczurek,
                          Phys.\ Rev.\ D {\bf 76}, 094014 (2007); 
                          A.~Rybarska, W.~Schafer and A.~Szczurek,
                          Phys.\ Lett.\ B {\bf 668}, 126 (2008);   
                          A.~Cisek, W.~Schafer and A.~Szczurek,
                          Phys.\ Rev.\ C {\bf 86}, 014905 (2012). 


\bibitem{vicmag_update}   V.~P.~Goncalves and M.~V.~T.~Machado, 
                          Phys.\ Rev.\ C {\bf 84}, 011902 (2011). 

\bibitem{gluon2}          A.~L.~Ayala Filho, V.~P.~Goncalves and M.~T.~Griep,
                          Phys.\ Rev.\ C {\bf 78},  044904 (2008); 
                          A.~Adeluyi and C.~Bertulani, 
                          Phys.\ Rev.\ C {\bf 84}, 024916 (2011); 
                          Phys.\ Rev.\ C {\bf 85}, 044904 (2012). 

\bibitem{motyka_watt}     L.~Motyka and G.~Watt, 
                          Phys.\ Rev.\ D {\bf 78}, 014023 (2008). 

\bibitem{Lappi}           T.~Lappi and H.~Mantysaari, 
                          Phys.\ Rev.\ C {\bf 87}, 032201 (2013). 


\bibitem{griep}    M.~B.~Gay Ducati, M.~T.~Griep and M.~V.~T.~Machado, 
                   Phys.\ Rev.\ D {\bf 88}, 017504 (2013); 
                   Phys.\ Rev.\ C {\bf 88}, 014910 (2013).
  
\bibitem{Guzey}    V.~Guzey and M.~Zhalov, 
                   JHEP {\bf 1310}, 207 (2013); JHEP {\bf 1402}, 046 (2014).

\bibitem{Martin}   S.~P.~Jones, A.~D.~Martin, M.~G.~Ryskin and T.~Teubner, 
                   JHEP {\bf 1311}, 085 (2013). 


\bibitem{glauber1}          G.~Sampaio dos Santos and M.V.T. ~Machado,
                            Phys.\ Rev.\ C {\bf 89}, 025201 (2014);                            Phys.\ Rev.\ C {\bf 91}, 025203 (2015).


\bibitem{bruno1}            V.~P.~Goncalves, B.~D.~Moreira and F.~S.~Navarra, 
                            Phys.\ Rev.\ C {\bf 90}, 015203 (2014);
                            Phys.\ Lett.\ B {\bf 742}, 172 (2015). 

\bibitem{Xie} 
Y.~p.~Xie and X.~Chen,
  Eur.\ Phys.\ J.\ C {\bf 76}, no. 6, 316 (2016); 
  Nucl.\ Phys.\ A {\bf 959}, 56 (2017).


\bibitem{bruno3}            V.~P.~Goncalves, B.~D.~Moreira and F.~S.~Navarra, 
                            Phys.\ Rev.\ D {\bf 95}, 054011 (2017).

\bibitem{vicnavdiego} 
  V.~P.~Goncalves, F.~S.~Navarra and D.~Spiering,
  Phys.\ Lett.\ B {\bf 768}, 299 (2017). 
  
\bibitem{tuchin} 
  G.~Chen, Y.~Li, P.~Maris, K.~Tuchin and J.~P.~Vary,
  Phys.\ Lett.\ B {\bf 769}, 477 (2017)  



\bibitem{cdf}      T.~Aaltonen {\it et al.}  [CDF Collaboration],
                   Phys.\ Rev.\ Lett.\  {\bf 102}, 242001 (2009).  

\bibitem{star}     C.~Adler {\it et al.}  [STAR Collaboration],
                   Phys.\ Rev.\ Lett.\  {\bf 89}, 272302 (2002). 
  
\bibitem{phenix}   S.~Afanasiev {\it et al.}  [PHENIX Collaboration],
                   Phys.\ Lett.\ B {\bf 679}, 321 (2009). 

\bibitem{alice}    B.~Abelev {\it et al.}  [ALICE Collaboration],
                   Phys.\ Lett.\ B {\bf 718}, 1273 (2013). 


\bibitem{alice2}   E.~Abbas {\it et al.}  [ALICE Collaboration],
                   Eur.\ Phys.\ J.\ C {\bf 73}, 2617 (2013). 
  
\bibitem{lhcb}              R. Aaij {\it et al.}  [LHCb Collaboration],
                            J.\ Phys.\ G {\bf 40}, 045001 (2013). 


\bibitem{lhcb2}             R. Aaij {\it et al.}  [LHCb Collaboration],
                            J.\ Phys.\ G {\bf 41}, 055002 (2014). 
   
\bibitem{lhcb3}             R.~Aaij {\it et al.} [LHCb Collaboration],
                            JHEP {\bf 1509}, 084 (2015). 
 
\bibitem{lhcbconf}          R.~Aaij {\it et al.} [LHCb Collaboration], 
                            LHCb-CONF-2016-007.

\bibitem{review_forward}    K.~Akiba {\it et al.} 
                            [LHC Forward Physics Working Group Collaboration], 
                            J.\ Phys.\ G {\bf 43}, 110201 (2016). 

\bibitem{wallon}            R.~Boussarie, A.~V.~Grabovsky, D.~Y.~Ivanov, 
                            L.~Szymanowski and S.~Wallon,
                            Phys.\ Rev.\ Lett.\  {\bf 119}, 072002 (2017).

  
\bibitem{Dress}             M.~Drees and D.~Zeppenfeld, 
                            Phys.\ Rev.\ D {\bf 39}, 2536 (1989).


\bibitem{nik}               N. N. Nikolaev, B. G. Zakharov,  
                            Phys. Lett. B  {\bf 332}, 184 (1994); 
                            Z. Phys. C {\bf 64}, 631 (1994).


\bibitem{vicmag_mesons}     V.~P.~Goncalves and M.~V.~T.~Machado,
                            Eur.\ Phys.\ J.\  C {\bf 38}, 319 (2004)

\bibitem{KMW}               H.~Kowalski, L.~Motyka and G.~Watt,  
                            Phys.\ Rev.\  D {\bf 74}, 074016 (2006). 
  
  

  
   


\bibitem{forshaw} 
  J.~R.~Forshaw and R.~Sandapen,
  Phys.\ Rev.\ Lett.\  {\bf 109}, 081601 (2012); M.~Ahmady, R.~Sandapen and N.~Sharma,
  Phys.\ Rev.\ D {\bf 94}, no. 7, 074018 (2016)
  
\bibitem{Li} 
  Y.~Li, P.~Maris, X.~Zhao and J.~P.~Vary,
  Phys.\ Lett.\ B {\bf 758}, 118 (2016)  


\bibitem{wflcg}
H.G. Dosch, T. Gousset, G. Kulzinger and H.J. Pirner,
Phys. Rev. {\bf D55}, 2602 (1997);\\
G. Kulzinger, H.G. Dosch and H.J. Pirner,
Eur. Phys. J. {\bf C7}, 73 (1999).

\bibitem{wfbg}
J.~Nemchik, N.~N.~Nikolaev, E.~Predazzi and B.~G.~Zakharov,
Z.\ Phys.\ C {\bf 75}, 71 (1997) 71.




\bibitem{sandapen}
J.~R.~Forshaw, R.~Sandapen and G.~Shaw,
  Phys.\ Rev.\ D {\bf 69}, 094013 (2004)
  

\bibitem{ipsat2}            H.~Kowalski and D.~Teaney, 
                            Phys.\ Rev.\ D {\bf 68}, 114005 (2003).

  

 \bibitem{armesto_amir}     N.~Armesto and A.~H.~Rezaeian,
                            Phys.\ Rev.\ D {\bf 90}, no. 5, 054003 (2014).

\bibitem{IIM_plb}           E.~Iancu, K.~Itakura and S.~Munier,
                            Phys.\ Lett.\ B {\bf 590}, 199 (2004).

\bibitem{Rezaeian_update}   A.~H.~Rezaeian and I.~Schmidt,
                            Phys.\ Rev.\ D {\bf 88}, 074016 (2013).

\bibitem{Watt_bcgc}         G.~Watt and H.~Kowalski, 
                            Phys.\ Rev.\ D {\bf 78}, 014016 (2008).

\bibitem{ipsat1}            J.~Bartels, K.~J.~Golec-Biernat and H.~Kowalski, 
                            Phys.\ Rev.\ D {\bf 66}, 014001 (2002).

\bibitem{ipsat3}            H.~Kowalski, T.~Lappi and R.~Venugopalan,  
                            Phys.\ Rev.\ Lett.\  {\bf 100}, 022303 (2008).

\bibitem{ipsat4}            A.~H.~Rezaeian, M.~Siddikov, M.~Van de Klundert and R.~Venugopalan,
                            Phys.\ Rev.\ D {\bf 87}, 034002 (2013).

\bibitem{armesto}           N.~Armesto,  Eur.\ Phys.\ J.\  C {\bf 26}, 35 (2002). 

  

\bibitem{erike}  
  E.~R.~Cazaroto, F.~Carvalho, V.~P.~Goncalves and F.~S.~Navarra,
  Phys.\ Lett.\ B {\bf 671}, 233 (2009)
  
  
\bibitem{raju} 
  H.~Kowalski, T.~Lappi and R.~Venugopalan,
  Phys.\ Rev.\ Lett.\  {\bf 100}, 022303 (2008);
  H.~Kowalski, T.~Lappi, C.~Marquet and R.~Venugopalan,
  Phys.\ Rev.\ C {\bf 78}, 045201 (2008).
  
\bibitem{simone} 
  M.~S.~Kugeratski, V.~P.~Goncalves and F.~S.~Navarra,
  Eur.\ Phys.\ J.\ C {\bf 46}, 465 (2006);
  Eur.\ Phys.\ J.\ C {\bf 46}, 413 (2006)  
  
\bibitem{bruno_double} 
V.~P.~Goncalves, B.~D.~Moreira and F.~S.~Navarra,
  Eur.\ Phys.\ J.\ C {\bf 76}, no. 3, 103 (2016); 
  Eur.\ Phys.\ J.\ C {\bf 76}, no. 7, 388 (2016).

\bibitem{diego_leading} 
  V.~P.~Goncalves, B.~D.~Moreira, F.~S.~Navarra and D.~Spiering,
  Phys.\ Rev.\ D {\bf 94}, no. 1, 014009 (2016)
  
  
\end{thebibliography}
\end{document}